\documentclass[twocolumn]{aastex701}

\usepackage{graphicx}
\usepackage{subfigure}
\usepackage{amsmath}
\usepackage{float}
\usepackage{placeins}
\usepackage{longtable}

\begin{document}
\shortauthors{Akhter et al.}
\title{Probing the Origin of X-ray Flares in the Low-Hard State of GRS 1915+105 Using \textit{AstroSat} and \textit{NuSTAR}}
\correspondingauthor{Shahzada Akhter}
\email{shahzadaakhter786phy@gmail.com}
\author[orcid=0009-0002-6849-0766,gname=Shahzada,sname=Akhter]{Shahzada Akhter}
\affiliation{Islamic University of Science and Technology, Awantipora 192122, Pulwama, India}
\email{shahzadaakhter786phy@gmail.com}

\author[orcid=0009-0001-5041-0927,gname=Sajad,sname=Boked]{Sajad Boked}
\affiliation{Department of Physics, University of Kashmir, Srinagar 190006, India}
\affiliation{Government Degree College, Bijbehara 192124, Anantnag, India}
\email{sboked@kashmiruniversity.ac.in} 

\author[orcid=0000-0002-0360-1851,gname=Bari,sname=Maqbool]{Bari Maqbool}
\affiliation{Islamic University of Science and Technology, Awantipora 192122, Pulwama, India}
\affiliation{Inter-University Centre for Astronomy and Astrophysics, Post Bag 4, Ganeshkhind, Pune 411007, India}
\email{barimaqbool@islamicuniversity.edu.in} 

\author[orcid=0000-0002-7609-2779,gname=Ranjeev,sname=Misra]{Ranjeev Misra}
\affiliation{Inter-University Centre for Astronomy and Astrophysics, Post Bag 4, Ganeshkhind, Pune 411007, India}
\email{rmisra@iucaa.in} 

\author[orcid=0000-0002-6449-9643,gname=V.,sname=Jithesh]{V. Jithesh}
\affiliation{Department of Physics and Electronics, Christ University, Hosur Main Road, Bengaluru - 560029, India}
\email{jithesh.v@christuniversity.in} 

\begin{abstract}
	We performed a detailed time-resolved spectral study of \textit{GRS 1915+105} during its low-flux rebrightening phase using the broadband capabilities of \textit{AstroSat} and \textit{NuSTAR} in May--June 2019. The \textit{AstroSat} light curves revealed erratic X-ray flares with count rates rising by a factor of $\sim$5. Flares with simultaneous LAXPC and SXT coverage were segmented and fitted using two degenerate but physically motivated spectral models: a reflection-dominated model (hereafter Model A) and an absorption-dominated model (hereafter Model B). In Model A, the inner disk radius (\( R_{\mathrm{in}} \)) shows a broken power-law dependence on flux, indicating rapid inward motion of the disk at higher flux levels. In contrast, Model B shows variable column density in the range of $10^{23}$ to $10^{24}$ cm$^{-2}$, displaying a strong anti-correlation with flux. Both models exhibit significant variation in the ionization parameter between low- and high-flux segments. The total unabsorbed luminosity in the 0.7--30~keV energy range ranged from $6.64 \times 10^{36}$ to $6.33 \times 10^{38}$~erg~s$^{-1}$. Across both models, several spectral parameters exhibited step-function-like behavior around flux thresholds of $5$--$10 \times 10^{-9}$ erg cm$^{-2}$ s$^{-1}$, indicating multiple spectral regimes. The disc flux contribution, more evident in Model B, increased with total flux, supporting an intrinsic origin for the variability. These findings point to a complex interplay between intrinsic disk emission, structured winds, and variable local absorption in driving the flare activity.
\end{abstract}

\section{Introduction} \label{sec:intro}
	
X-ray binaries (XRBs) are astrophysical systems consisting of a compact object and a companion star, emitting high-energy radiation. They are classified based on the mass of companion star into Low Mass X-ray Binaries (LMXBs) and High Mass X-ray Binaries (HMXBs) \citep{Bradt1983}. LMXBs have companion stars with masses below $1M_{\odot}$, usually accreting matter via Roche lobe overflow, whereas HMXBs feature companions with masses exceeding $10M_{\odot}$, primarily accreting through stellar winds. Black hole X-ray binaries (BHXRBs) exhibit distinct X-ray emission states, reflecting different accretion configurations. In the soft state, X-rays primarily originate from the thermal radiation of the accretion disk \citep{Shakura1973}, while in the hard state they result from interactions between photons and high-energy electrons in the hot corona \citep{Shakura1976, Sunyaev1980, Lightman1987}. Observations suggest that the corona is significantly hotter than the accretion disk, with electron temperatures reaching around 100 keV, compared to the disk temperature of approximately 0.1 keV. The exact shape and position of the corona relative to the disk remain uncertain, and different geometries such as the lamppost corona, sandwich corona, etc. have been proposed \citep{haardt1991,stern1995,zdziarski1998}. When X-rays from the corona are reflected off the accretion disk, they produce characteristic emission features, including the iron K$\alpha$ fluorescence line at 6.4 keV \citep{George1991, Ross2005}. This reflection spectrum provides crucial insights into the velocity of the orbiting disk gas and reveals both special and general relativistic effects near the black hole.

GRS 1915+105, a superluminal XRB, was first observed in outburst by {\it WATCH} all-sky monitor on board the {\it GRANAT} satellite, in August 1992 \citep{CastroTirado1992}. GRS 1915+105 hosts a black hole with a mass of $12.5^{+2.0}_{-1.8} \, M_{\odot}$, situated at a radio parallax distance of $8.6^{+2.0}_{-1.6}$ kpc and the accretion disk is inclined at an angle of $60^\circ$ \citep{Reid2014}. Unlike typical LMXBs, which spend long periods in quiescence before undergoing outbursts lasting from months to a few years, GRS 1915+105 exhibited persistent brightness from its discovery until mid-2018, after which its X-ray flux began an exponential decline, signaling a major shift in its long-term behavior. This decline marked the lowest soft X-ray flux recorded for the system in 22 years of continuous monitoring by MAXI/GSC and RXTE/ASM \citep{Negoro2018}. The reduction in flux was interpreted as the system entering a quiescent phase, similar to transitions observed in other black hole X-ray binaries (BH-XRBs). However, in May 2019 (around MJD 58600), GRS 1915+105 exhibited a resurgence in activity, despite maintaining lower-than-average X-ray flux levels, with radio flares during this period suggesting high mass accretion rates \citep{MOTTA2019,koljonen2021}. \citep{Zhou2025} reported a significant reduction in ionization($\log \xi$)  of accretion disk winds during this transition into the obscured state. During this dim state, GRS 1915+105 displayed behavior distinct from its historical trends, characterized by intense and irregular X-ray flares \citep{Iwakiri2019, Jithesh2019, Neilsen2019}. Observations revealed phenomena such as Compton-thick obscuration, hard X-ray spectra, prominent emission and absorption lines, and significantly increased intrinsic absorption, with column densities exceeding those typically observed in the source \citep{Motta2021, Balakrishnan2021, Koljonen2020, Miller2019a, Miller2019b, Miller2020}.

The reflection spectrum during low flux periods indicated unusually high reflection fractions (greater than unity), reminiscent of behaviors observed low flux state of active galactic nuclei (AGN) such as MCG-6-30-15 \citep{Fabian2003}, 1H 0707-495 \citep{Fabian2004}, and 1H 0419-577 \citep{Fabian2005}. \citet{Miniutti2004} proposed the light-bending model to explain such high reflection fractions. In this model, X-ray photons emitted from regions close to the black hole are either absorbed by the black hole itself or illuminate the inner region of accretion disk. This concentration amplifies the reflection-dominated spectral component due to intense light-bending effects near the black hole. They identified three distinct flux states low, moderate, and high, where the reflection-dominated component varies in its relationship with the direct continuum, from correlation to anticorrelation and weak interaction. Similar reflection-dominated spectra have been observed in other BH-XRBs for instance, studies of RXTE data for XTE J1650-500 by \citet{Rossi2005} and \citet{Reis2013} revealed an increase in reflection fraction during transitions between soft and hard intermediate states. These analyses also demonstrated an anticorrelation between power-law flux and line equivalent width, alongside a positive correlation between power-law flux and Fe K$\alpha$ line flux at low power-law flux levels, aligning with the predictions of the light-bending model.

\cite{Koljonen2020} conducted a detailed spectral analysis of several black hole X-ray binaries using \textit{NuSTAR} observations. For GRS 1915+105 in particular, they analyzed three observational epochs using a complex combination of spectral components, including smeared absorption edges, partial covering, distant and relativistic reflection, and emission or obscuration from surrounding tori. The first epoch was adequately described by either a relativistic reflection model or a torus scenario, while the latter two epochs required models with strong partial covering featuring high column densities, or alternatively, a torus with radial motion. \cite{Sajad2024} conducted time-resolved spectroscopy of a flare from the same source using \textit{AstroSat}, employing degenerate spectral models, considering reflection and absorption dominated scenarios. Their spectral modeling included thermal Comptonization, multicolor disk emission, photoionized absorption, and relativistic reflection components. In both interpretations, the flaring was attributed to intrinsic variability, with the reflection-dominated case resembling low flux AGN spectra.

The {\it AstroSat} observations analyzed in this study were carried out during the onset of the rebrightening phase of the source, as detected by {\it MAXI/GSC} \citep{Iwakiri2019}. As shown in Figure~\ref{Maxi_lc}, the dashed line marks the time of one such {\it AstroSat} observation, during which the one-day binned {\it MAXI} count rate indicates signs of rebrightening. All {\it AstroSat} observations that captured flaring activity were included in this analysis. Prominent X-ray flares were observed in both the Large Area X-ray Proportional Counter (LAXPC) and the Soft X-ray Telescope (SXT) instruments onboard {\it AstroSat}, with the source intensity increasing by nearly a factor of five. These strong flares gradually weakened after the first 60 ks of the observation and eventually disappeared. A similar, though comparatively weaker, episode of flaring activity was again detected in both instruments during the {\it AstroSat} observation conducted in June 2019. \citet{Sajad2024} previously conducted time-resolved spectroscopy of a single flare from the May 2019 observation using a degenerate spectral modeling approach. In this study, we extend their analysis by incorporating all flares observed by {\it AstroSat} where simultaneous flaring was detected in both LAXPC and SXT. One of the \textit{NuSTAR} observations [\citep{Koljonen2020} Epoch 2 data which was conducted just a few days after our May-2019 {\it AstroSat} observation was also re-analyzed using the same spectral models to verify our results.

This paper is organized as follows: In Section \ref{sec:datareduction} , we provide a brief overview of the data reduction procedures for the observations collected from the SXT and LAXPC instruments onboard \textit{AstroSat}, as well as from\textit{NuSTAR}. Section \ref{sec:spectral} presents the results of our spectral analysis for the time-resolved data. We then discuss and interpret these findings in Section \ref{sec:discussion}. 

\section{OBSERVATION AND DATA REDUCTION} \label{sec:datareduction}

\begin{figure*}[!htbp]
		\centering
		\includegraphics[width=2.0\columnwidth]{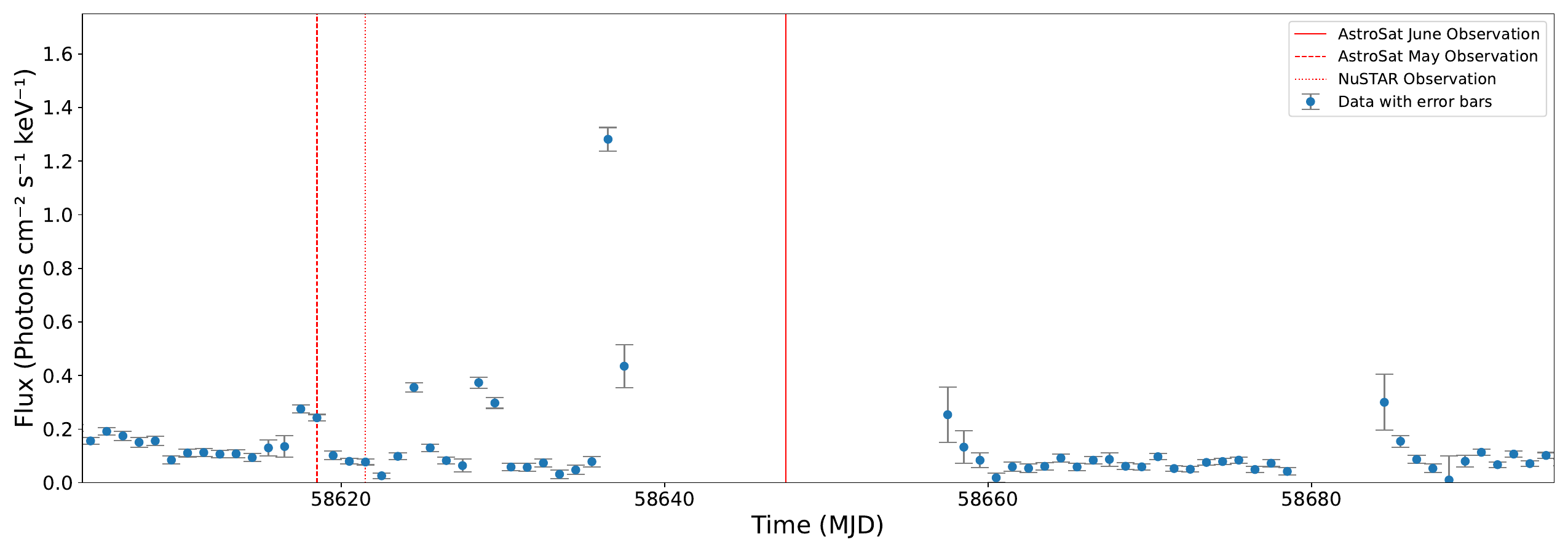}
		\caption {1-day binned {\it MAXI/GSC} light curve of GRS 1915+105 from MJD 58604 to 58694. 
			The data points with error bars represent the daily averaged flux values with their corresponding uncertainties. 
			The dashed, dotted, and solid vertical lines indicate the times of our May \textit{AstroSat} observation, \textit{NuSTAR} observation, and June \textit{AstroSat} observation, respectively (see Table~\ref{obs_log}).}
		\label{Maxi_lc}
	\end{figure*}

In this work, we analyzed \textit{AstroSat} and \textit{NuSTAR} observations of GRS 1915+105 during May and June 2019. Table~\ref{obs_log} provides a summary of these observations.  These data are overlaid on the 1-day binned 2–20 keV \textit{MAXI} lightcurve shown in Figure~\ref{Maxi_lc}, highlighting the observation times of both {\it AstroSat} and {\it NuSTAR}.

	
	
	\begin{table*}
		\caption{Observation log.}
		\label{obs_log}
		\centering
		\begin{tabular}{cccccc}
			\hline
			\textbf{Instrument}&	\textbf{Obs ID} & \textbf{Start Time} & \textbf{End Time} & \textbf{MJD} & \textbf{Exposure} \\
			&&\textbf{Date}&\textbf{Date}&&\textbf{ks} \\
			\hline
			{\it AstroSat}&	T03-116T01-9000002916 &  02:10:37 & 07:14:38 & 58618.090 & 39.99 \\
			&&15-05-2019&16-05-2019& \\
			
			&	T03-117T01-9000002988 &  13:48:46 &  13:59:42 & 58647.575 &35.44 \\
			&&13-06-2019&14-06-2019& \\
			\hline
			{\it NuSTAR}&	30502008002 & 12:41:49 &  03:19:35 & 58622.52845 &25.40 \\
			&&19-05-2019&20-05-2019& \\
			\hline
		\end{tabular}
	\end{table*}

\subsection{AstroSat}

In this work, we analyzed \textit{AstroSat} observations of GRS 1915+105 during May and June 2019, where the flaring was seen during the low state of the source. The SXT Level 2 data was obtained from the {\it AstroSat data archive-ISSDC} in photon-counting mode. The data reduction tools used were provided by the SXT POC team at TIFR. To merge individual clean event files for each orbit into a single merged clean event file, we used \texttt{SXTPYJULIAMERGER V01}. These merged clean event files were then used to extract images, light curves, and spectra using \texttt{XSELECT} from \texttt{HEASOFT V6.31}. Corresponding auxiliary response files (ARFs) were generated using the \texttt{sxtARFModule}, while the response matrix file (\texttt{sxt\_pc\_mat\_g0to4.rmf}) and blank sky background spectrum file (\texttt{SkyBkg\_comb\_EL3p5\_Cl\_Rd16p0\_v01.pha}) provided by the SXT instrument team were used for spectral analysis. Circular source regions with a radius of 18 arcmin were extracted using \texttt{ds9}, centered on the source. To account for the pileup effect, we extracted the annulus source region with an outer radius of 18 arcmin and an inner radius of 2.5 arcmin for segments, where the net count rate was greater than 40 counts per second. Afterwards, using the interactive command {\sc grppha}, the background spectrum, vignetting-corrected ARF, and RMF were combined so that each bin included a minimum of 25 counts.

The \textit{LAXPC}, one of the primary instruments onboard \textit{AstroSat}, comprises three identical co-aligned X-ray proportional counter units \textit{LAXPC10}, \textit{LAXPC20}, and \textit{LAXPC30}. It offers high time resolution of 10 µs and covers wide energy range of 3.0–80.0 keV \citep{Yadav2016, Antia2017, Agarwal2006}. Because \textit{LAXPC10} and \textit{LAXPC30} were not operational at the time of our observation, we used only the data from \textit{LAXPC20}. The Level-1 \textit{LAXPC} data were obtained from the \textit{AstroSat} data archive hosted by \textit{ISSDC} and processed using the standard \texttt{LAXPCsoft} pipeline. The Level-2 event file was then created using \texttt{laxpc\_make\_event}, and Good Time Intervals (GTIs) were derived using \texttt{laxpc\_make\_stdgti} with the housekeeping (filter) files. Light curves and spectra were extracted from the Level-2 event file using \texttt{laxpc\_make\_lightcurve} and \texttt{laxpc\_make\_spectra}, respectively. Instrumental background was estimated by generating background spectra and light curves with \texttt{laxpc\_make\_backspectra} and \texttt{laxpc\_make\_backlightcurve}, using the same GTIs and energy calibration files. This procedure yielded background-subtracted light curves and calibrated spectra suitable for detailed timing and spectral analysis.

\begin{figure*}
	      \includegraphics[scale=0.35]{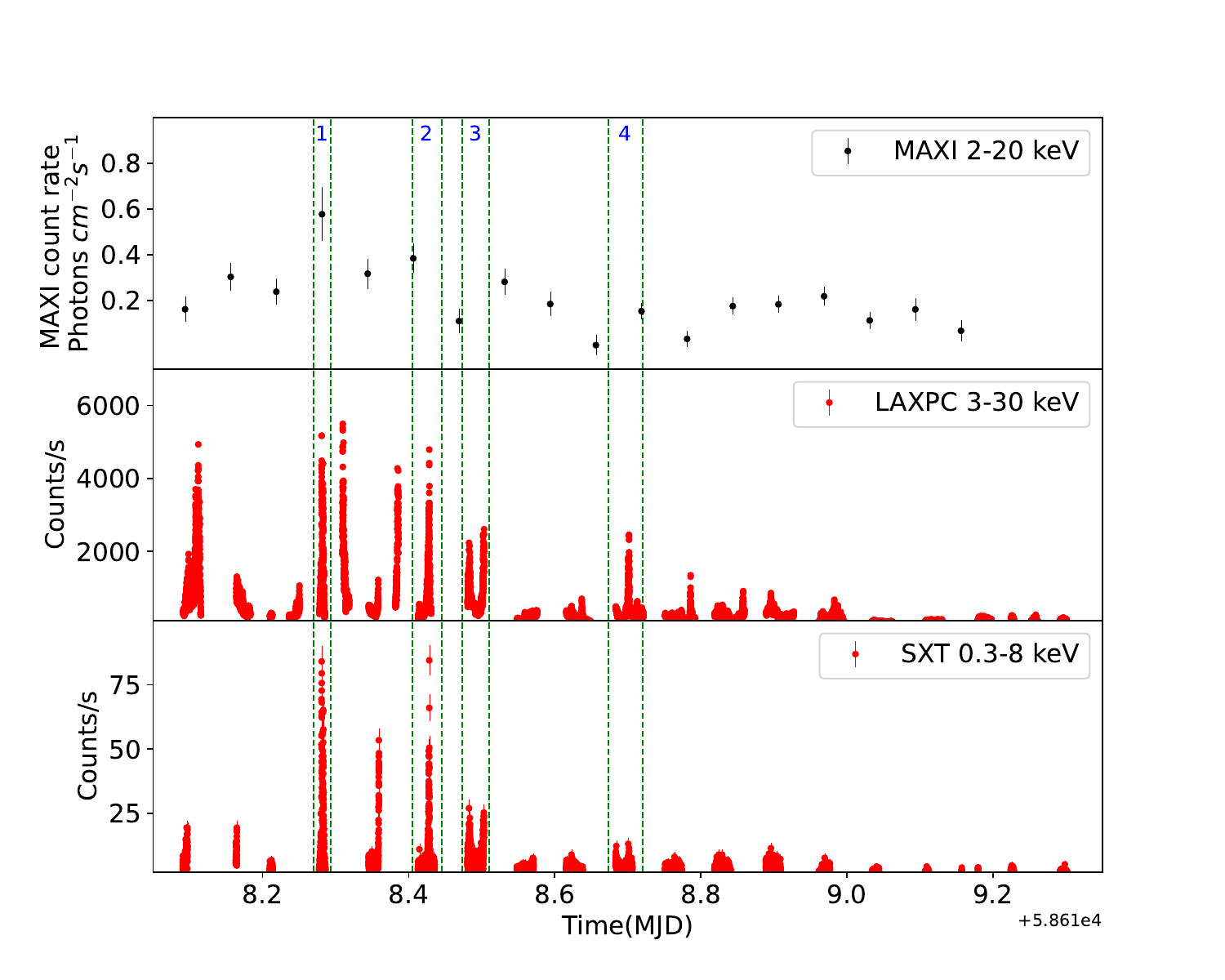}
              \includegraphics[scale=0.35]{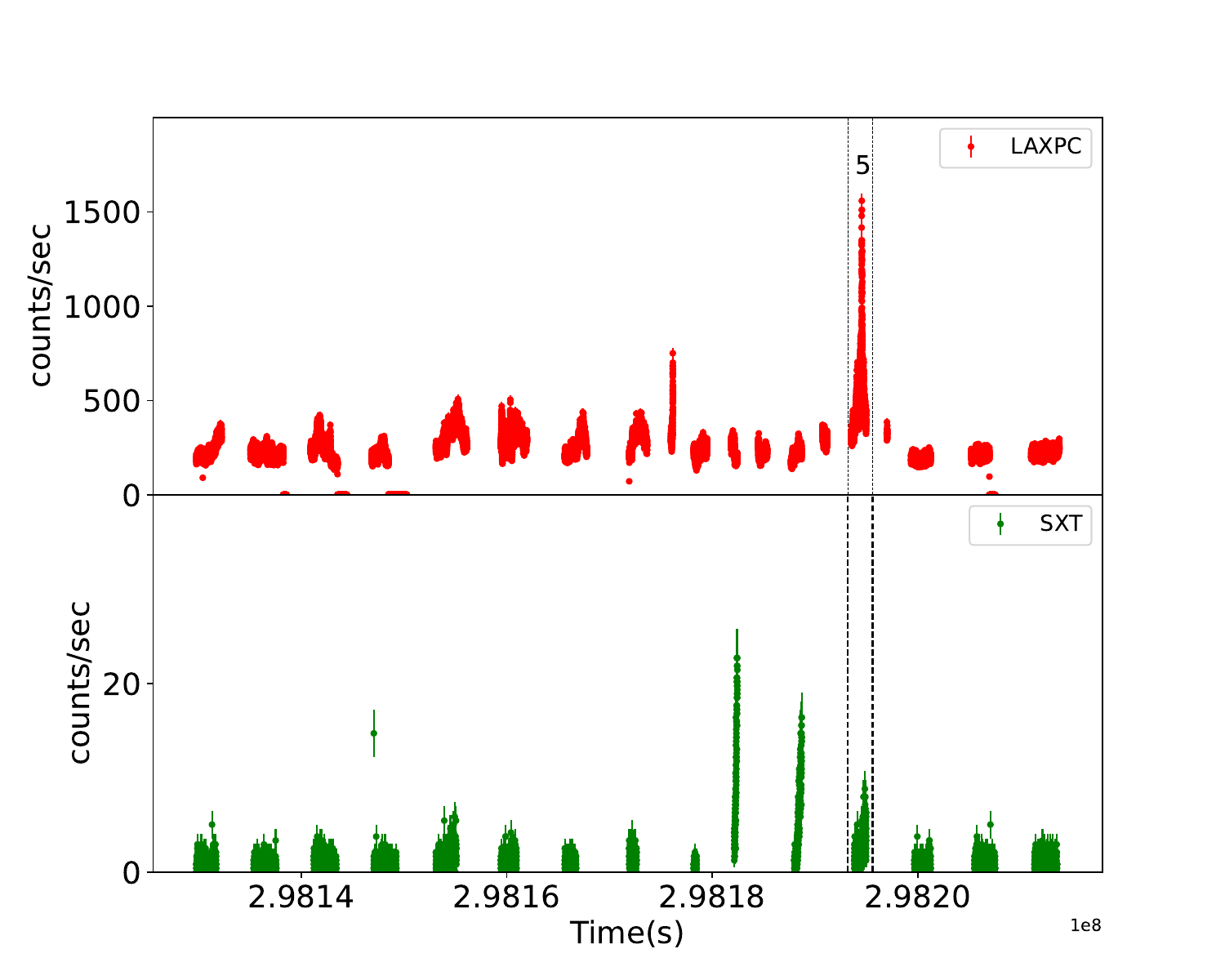}
	\caption{Left:\textit{MAXI} and \textit{AstroSat} lightcurves for May observation:Top panel shows 1.5-hour binned {\it MAXI} lightcurve in the energy range of 2.0--20.0 keV for the \textit{AstroSat} observation period; the middle panel shows the background-subtracted LAXPC lightcurve in the 3.0--30.0 keV energy band with a binning of 2.3775 s; and the bottom panel displays the SXT lightcurve in the 0.3--8.0 keV energy band also binned at 2.3775 s. Right: \textit{AstroSat} lightcurves for June observation:The top panel shows the background-subtracted LAXPC lightcurve in the 3.0--30.0 keV energy band; and the bottom panel displays the SXT lightcurve in the 0.3--8.0 keV energy band both with 2.3775 s time resolution. The flares marked between dotted vertical lines, labelled 1 to 5 were used for this analysis.}
	
	\label{june_full_lightcurve}
	\end{figure*}

\begin{figure*}
    \centering
     \includegraphics[scale=0.35]{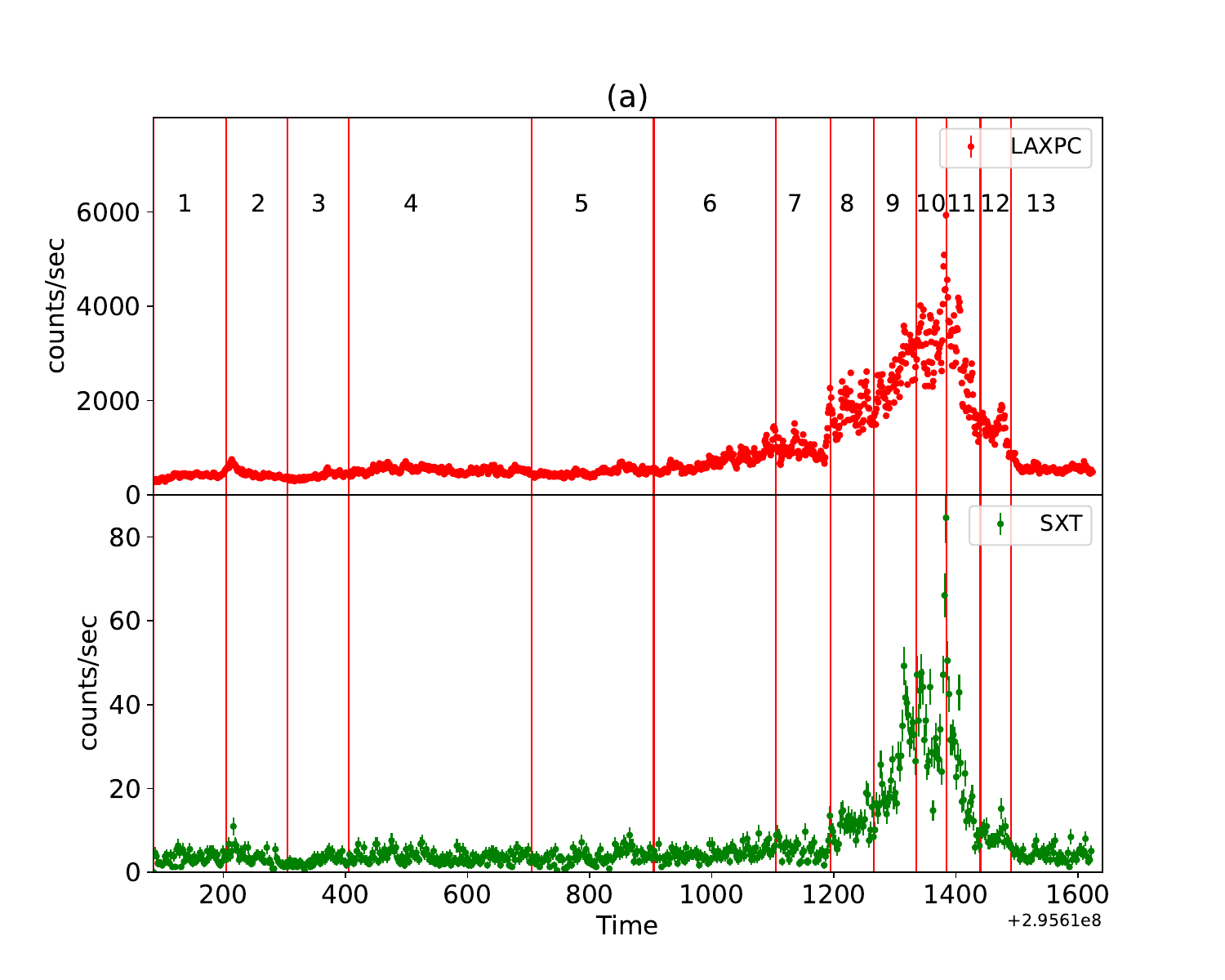}
     \includegraphics[scale=0.35]{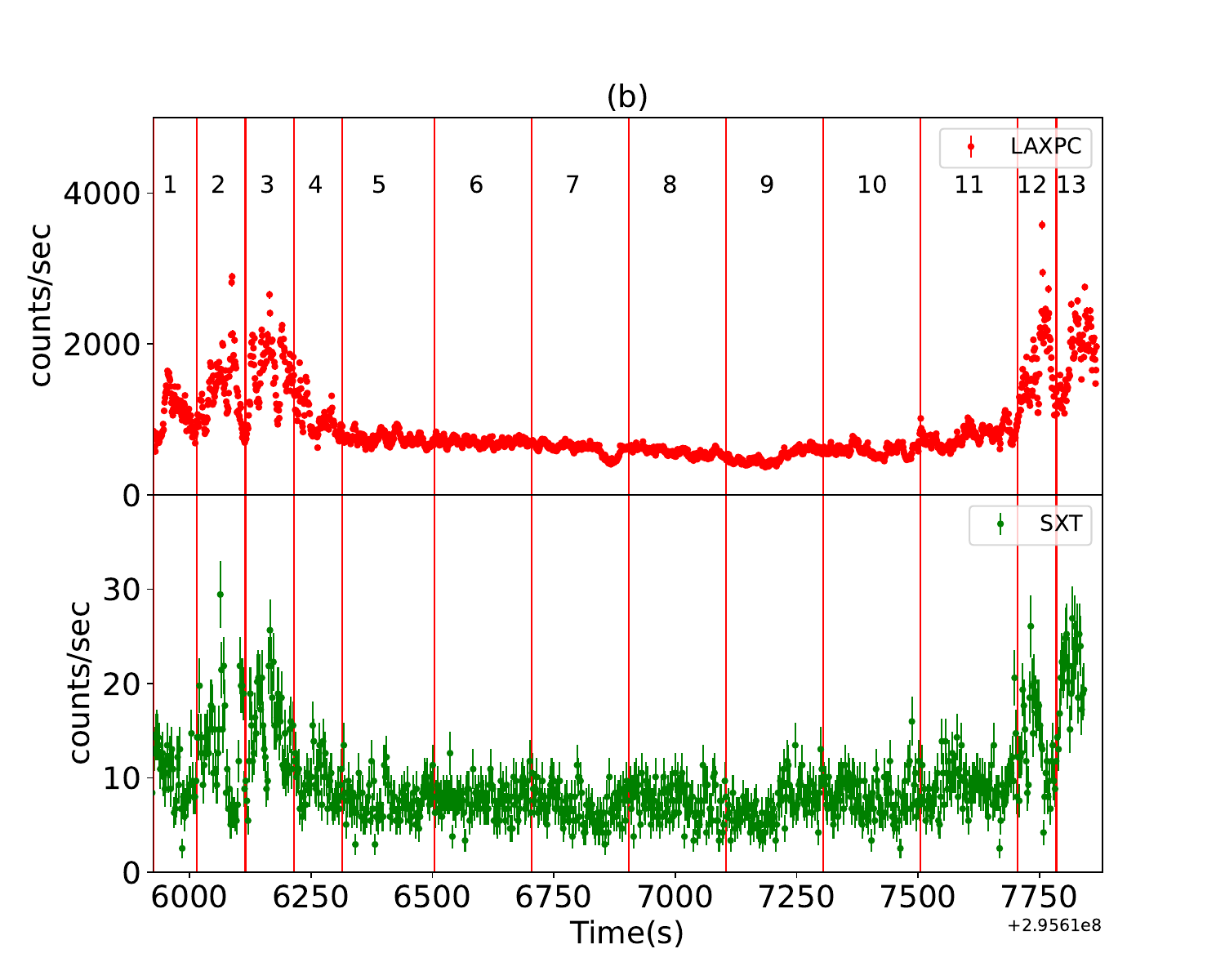}
     \includegraphics[scale=0.35]{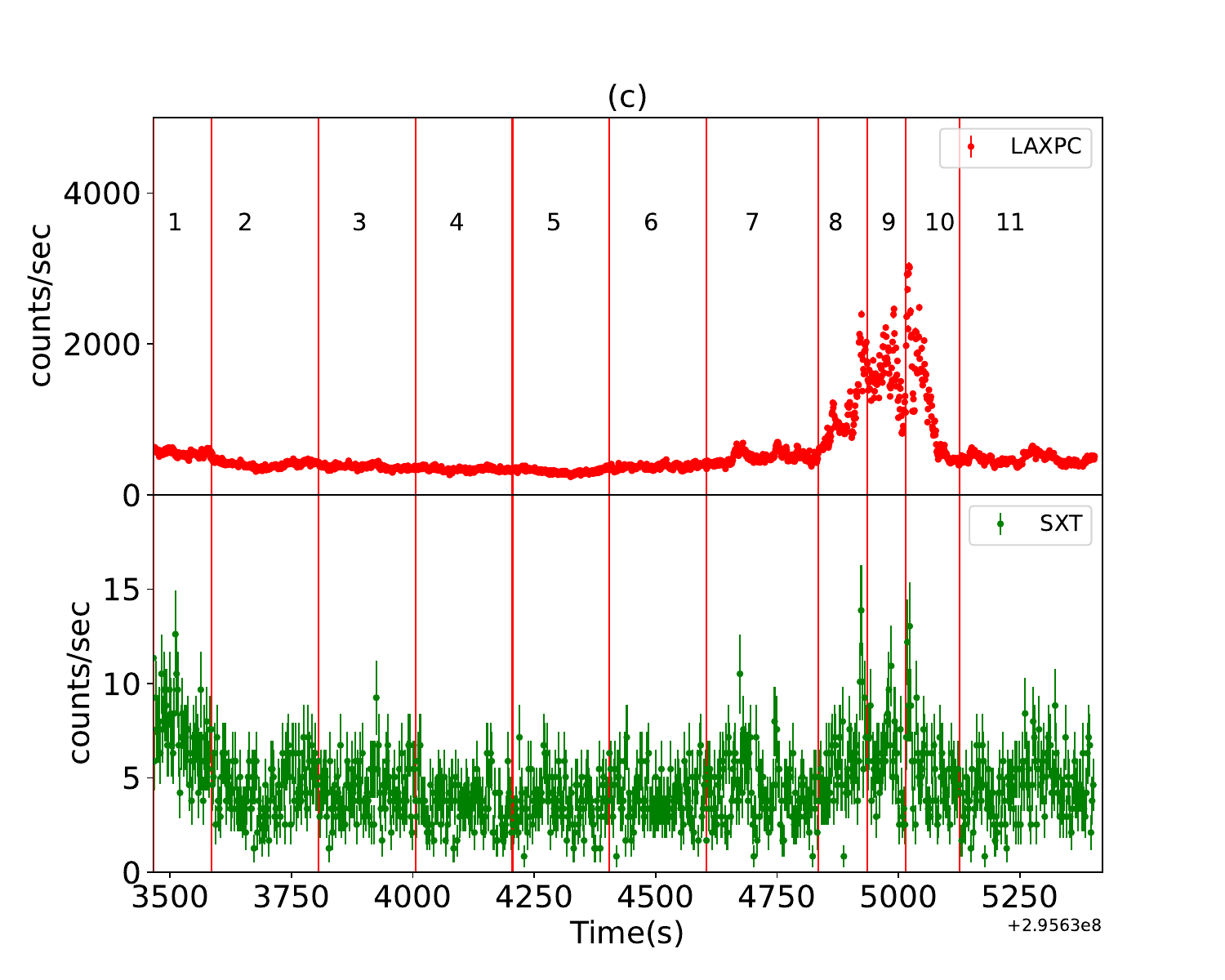}
     \includegraphics[scale=0.35]{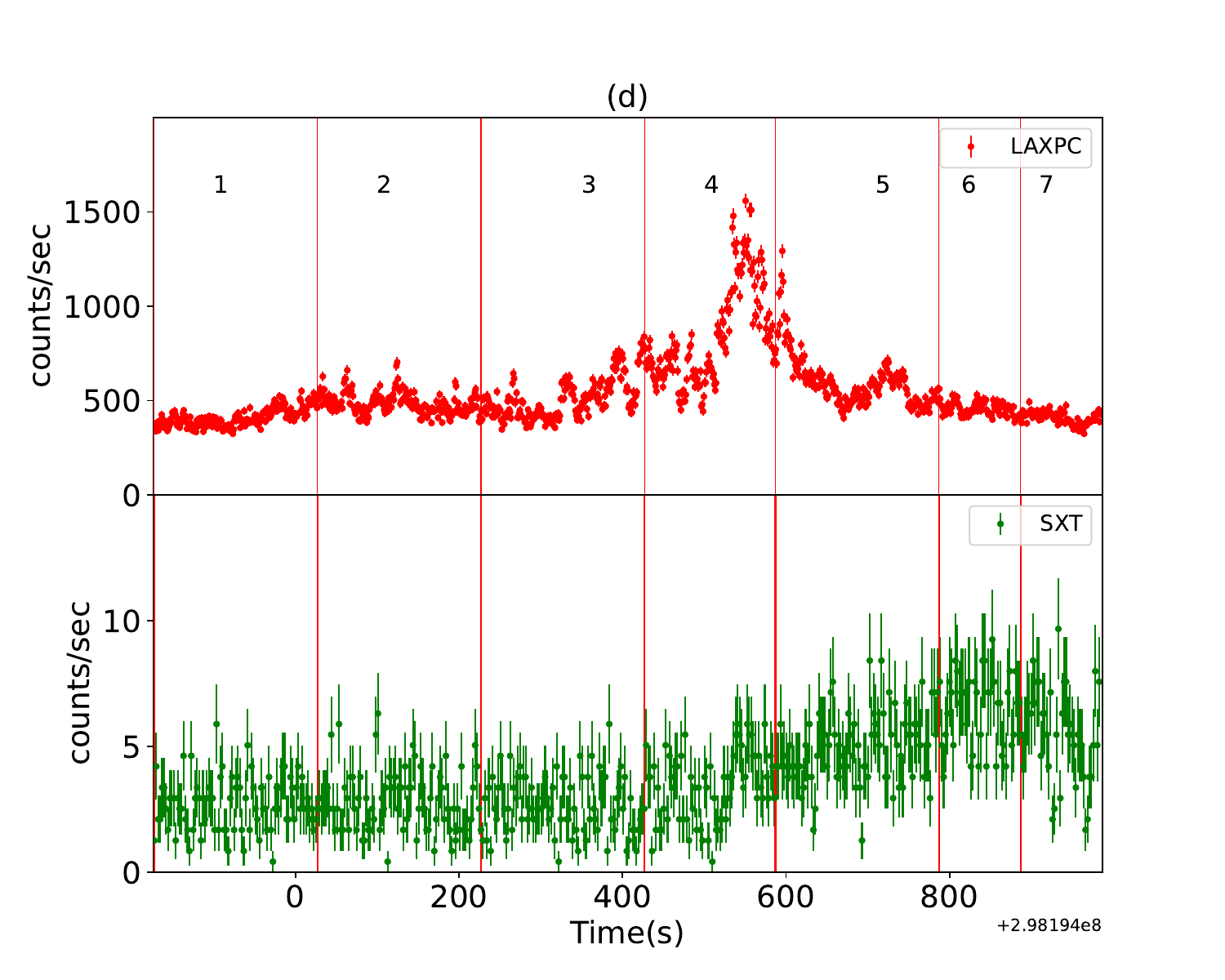}
		\caption{Figures (a)-(d) show zoomed-in views of Flares 2, 3, 4, and 5, marked in Figure~\ref{june_full_lightcurve}, with top and bottom panels showing {\it LAXPC} and {\it SXT} lightcurves,respectively, both with 2.3775 s time resolution. In each panel, the X-axis represents the time elapsed since the onset of the corresponding flare in \textit{AstroSat} seconds. Each flare is divided into multiple segments, labeled numerically and separated by vertical lines.}
		
\label{All_flares}
\end{figure*}

In this work, we have analyzed all the flares marked by dotted lines, numbered 1 to 5 (hereafter flare1, flare2, flare3, flare4 and flare5) in Figure~\ref{june_full_lightcurve}, each of which had simultaneous data availiable in both the \textit{LAXPC} and \textit{SXT} instruments. Flare 1 was previously analyzed by \cite{Sajad2024} (Figure 4 therein) by dividing it into 12 segments. Each of the flares (2, 3, 4, and 5) was divided into multiple segments, as shown in Figure~\ref{All_flares}, with the segments labeled numerically. For each segment across all flares, a single Good Time Interval (GTI) was used to extract spectra and light curves from both instruments.

\subsection{NuSTAR}

The Nuclear Spectroscopic Telescope Array (\textit{NuSTAR}) is the first orbiting telescope capable of focusing hard X-rays \citep{Harrison2013}. It operates in the 3–79 keV energy range, significantly extending the sensitivity of focusing optics beyond the ~10 keV limit achieved by earlier X-ray satellites. \textit{NuSTAR} consists of two identical, co-aligned X-ray telescopes that focus hard X-rays across a broad energy range of 3–79 keV. Each telescope has its own focal plane modules, \textit{FPMA} and \textit{FPMB} \citep{Harrison2013}. For this study, we used \textit{NuSTAR} observations of GRS 1915+105 conducted on May 19-20, 2019 (Observation ID: 30502008002). The raw data were processed following the \textit{NuSTAR} Data Analysis Software Guide using the \textit{NuSTAR} data analysis software (\textit{\footnotesize NUSTARDAS V2.1.1)} provided with {\it HEASOFT V6.31} and \textit{CALDB} version 20221229. Clean, filtered event files were generated using the \textit{NUPIPELINE} routine with the parameters TENTACLE=YES and SAAMODE=OPTIMISED to remove intervals with high background. Science products for both \textit{FPMA} and \textit{FPMB} were extracted using the \textit{NUPRODUCTS} routine. The source spectrum was extracted using a circular region with a radius of 100 arcseconds centered on the source, while the background spectrum was taken from a 100 arcsecond circular region in a detector area free of source contamination. Finally, the data from both detectors were grouped such that each bin contains a minimum of 30 counts, and the corresponding background and ancillary response files were used for the analysis..

	\begin{figure} 
		\centering
		\includegraphics[width=0.96\columnwidth]{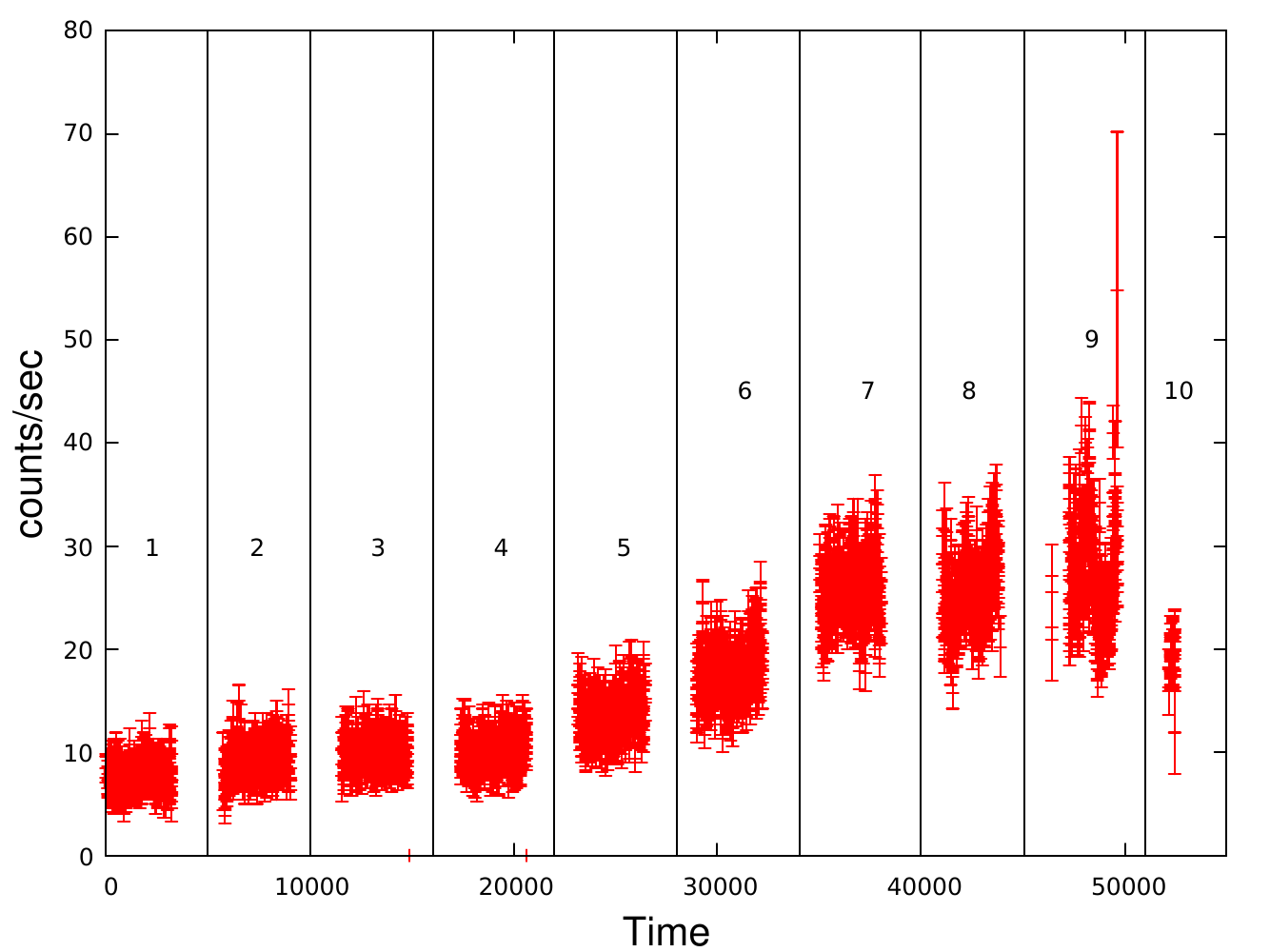}
		\caption{{\it NuSTAR} lightcurve of the observation used for this study. The time along X-axis is the time since 58622.52845 MJD (Start time of observation). The lightcurve is obtained by FPMA detector and binned at 10 s. we have divided the lightcurve into 10 segments,labelled 1 to 10 each marked by vertical lines}
		\label{nustar_lc}
	\end{figure}

\section{Spectral Analysis} \label{sec:spectral}

All spectral fits were performed using \textit{XSPEC version 12.12.0}  \citep{Arnaud1996}. To account for cross-calibration uncertainties between different instruments, a constant factor was included in all joint spectral fits. For \textit{AstroSat}, the constant was fixed at unity for \textit{SXT} and allowed to vary for \textit{LAXPC}. Similarly, for \textit{NuSTAR}, the constant was fixed at unity for \textit{FPMA} and allowed to vary for \textit{FPMB}. Additionally, a gain fit was applied to both \textit{LAXPC} and \textit{SXT} to adjust the gain of the response file \citep{Antia2021}. A 3\% systematic error was included in the joint fitting of \textit{AstroSat} \textit{SXT} and \textit{LAXPC} spectra to account for uncertainties in the response matrix. All parameter errors are reported at the 90\% confidence level.

The \textit{NuSTAR} observation(ObsID: 30502008002) used in this study was previously analyzed by \cite{Koljonen2020}. They modeled the spectrum assuming either complete reprocessing of the intrinsic emission by surrounding neutral, partially covered material or partial obscuration by a thick torus with a variable opening angle. Their model (Model A) included a neutral partial covering absorber \textit{pcfabs} along with two \textit{xillverCp} components to represent reflection from regions with differing ionization. Their analysis revealed that the emission is dominated by a hard Comptonized continuum ($\Gamma = 1.58$, $kT_e = 15.6\,\text{keV}$) reprocessed in a stratified reflecting medium with both highly and mildly ionized layers ($\log \xi = 3.4$ and $\log \xi = 2.0$, respectively). The spectrum also showed evidence of partial covering absorption with $N_H \sim 4.1 \times 10^{23}\,\text{cm}^{-2}$ and a covering fraction of $\sim 0.58$, suggesting inhomogeneous material partially obscuring the source.

Like the typical hard state spectra of BH-XRBs, we first modeled the spectrum with a Comptonization component (thcomp; \cite{Zdziarski2020}), a multicoloured black-body component (diskbb; \cite{Mitsuda1984}), and Galactic absorption (tbabs; \cite{Wilms2000}), along with a Gaussian to counter the residuals from the Fe K$\alpha$ emission line at 6.4 keV. The total model in XSPEC notation was $\text{constant} \times \text{tbabs} (\text{thcomp} \times \text{diskbb} + \text{gaussian})$. This model provided a good fit for all segments; however, the electron temperature of the Comptonization component turned out to be very low with high optical depth, causing the component's peak to occur around 20 keV. At this high optical depth and low electron temperatures, the thermal electrons cannot efficiently upscatter low-energy photons to high energies, leading the Comptonized component to fit more of a reflection-like feature, including the Compton hump at around 20 keV.

\subsection{Spectral Parameter Trends with Flux Using Reflection Dominated Model} \label{sec:spectral_reflection}

The \texttt{relxillCp} flavour of \texttt{relxill} family of models is commonly employed in the spectral study of BH-XRBs \citep{You2021}. It combines the reflection model \texttt{xillver} \citep{GarciaKallman2010, GarciaKallmanMushotzky2011, Garcia2013} with the relativistic model \texttt{relline} \citep{Dauser2010, Dauser2013}. It incorporates a thermally Comptonized continuum (\texttt{nthcomp}; \citep{Zdziarski1996,Zycki1999}) as the illuminating source. For {\it AstroSat} we fitted the joint {\it SXT} and {\it LAXPC} spectra in the energy ranges of 0.7-7 and 3-30 keV respectively, and for {\it NuSTAR} the joint {\it FPMA} and {\it FPMB} spectra was fitted in the energy range of 3-79 keV. The model \texttt{TBabs} was employed to account for Galactic absorption, paired with the relativistic reflection model \texttt{relxillCp} to fit both our \textit{AstroSat} and \textit{NuSTAR} spectra. In \textit{XSPEC} notation, the total model is expressed as \(\texttt{constant} \times \texttt{TBabs} \times \texttt{relxillCp}\) (hereafter Model A).

\begin{table*}
	\centering
	\caption{The Spearman's correlation coefficients and p-values, along with the power-law fit parameters (\( f(x) = A \times x^{-p} \)) for the relationship between model parameters and flux using Model A.}

\begin{tabular}{llcccc}
	\hline
	\textbf{Parameter} & \textbf{A} & \textbf{p} & \textbf{ Correlation coefficient} & \textbf{P-value}   \\ 
	\hline
	relxillCp ($\Gamma$)   & $1.90_{-0.03}^{+0.03}$ & $-0.02_{-0.01}^{+0.01}$ & 0.40 & $ 2.07 \times 10^{-4}$  \\ 
	TBabs $N_\mathrm{H}$          & $3.91_{-0.25}^{+0.25}$ & $-0.18_{-0.02}^{+0.02}$ & 0.73 & $5.5 \times 10^{-7}$  \\  
	relxillCp $\log(\xi)$   & $1.93_{-0.12}^{+0.12}$ & $-0.13_{-0.03}^{+0.03}$  & 0.34 & $2.14 \times 10^{-3}$  \\   
	relxillCp $R_\mathrm{in}$  &  $1.81_{-0.04}^{+0.04}$&$0.10_{-0.01}^{+0.01}$&-0.89&$1.5 \times 10^{-13}$ \\  
	\hline
\end{tabular}
\label{Table 2}
\end{table*}

High reflection fraction has been previously reported for the source, so we modeled \texttt{relxillCp} purely as a reflection component by fixing \texttt{refl\_frac} to -1 (following \cite{Sajad2024}). Due to the strong degeneracy between the black hole spin and the inner disc radius, we fixed the spin parameter to $a = 0.998$, assuming a maximally spinning black hole \citep{McClintock2006}. The inner disc radius ($R_{\text{in}}$) was left free, while the outer radius was fixed at $R_{\text{out}} = 400\,R_g$, where $R_g = GM/c^2$ is the gravitational radius. The inclination angle was fixed at $60^\circ$, based on \citet{Reid2014}. The emissivity profile in the \texttt{relxillCp} model follows $r \propto r^{-q_{\text{in}}}$ for $r < R_{\text{br}}$ and $r \propto r^{-q_{\text{out}}}$ for $r > R_{\text{br}}$, with $R_{\text{br}}$ as the break radius. We assumed a single power-law emissivity profile ($q_{\text{in}} = q_{\text{out}}$) and allowed it to vary freely.
The \texttt{relxillCp} model allows the disc density to vary between $10^{15}$ to $10^{20}$\,cm$^{-3}$. From fitting all the segments, the average disc density was found to be around $10^{17}$\,cm$^{-3}$. As variations in this parameter did not significantly affect the fit statistics, we fixed $\log N$ to 17\,cm$^{-3}$ in subsequent fits. The iron abundance ($A_{\mathrm{Fe}}$) was fixed at solar value. The ionization parameter in the \texttt{relxill} model is defined as the ratio of the incident X-ray flux to the gas density, from $\log \xi = 0$ (neutral) to $\log \xi = 4.7$ (highly ionized). 
We fixed the electron temperature of the corona (\(kT_e\)) at 400\,keV in the \texttt{relxillCp} model (following \cite{Sajad2024}). 
The unabsorbed flux was estimated using the \texttt{cflux} convolution model in \textit{XSPEC}. To examine the relationship between the model parameters and the unabsorbed flux, we used Spearman's  correlation and fitted the data with a power-law of the form:(\( f(x) = A \times x^{-p} \)) where $A$ is the normalization factor and $p$ is the power-law index. The Spearman's correlation coefficients, corresponding p-values, and the power-law fit parameters are presented in Table~\ref{Table 2}.

\begin{figure*}[!htbp]
		
		     \includegraphics[scale=0.35,angle=-90]{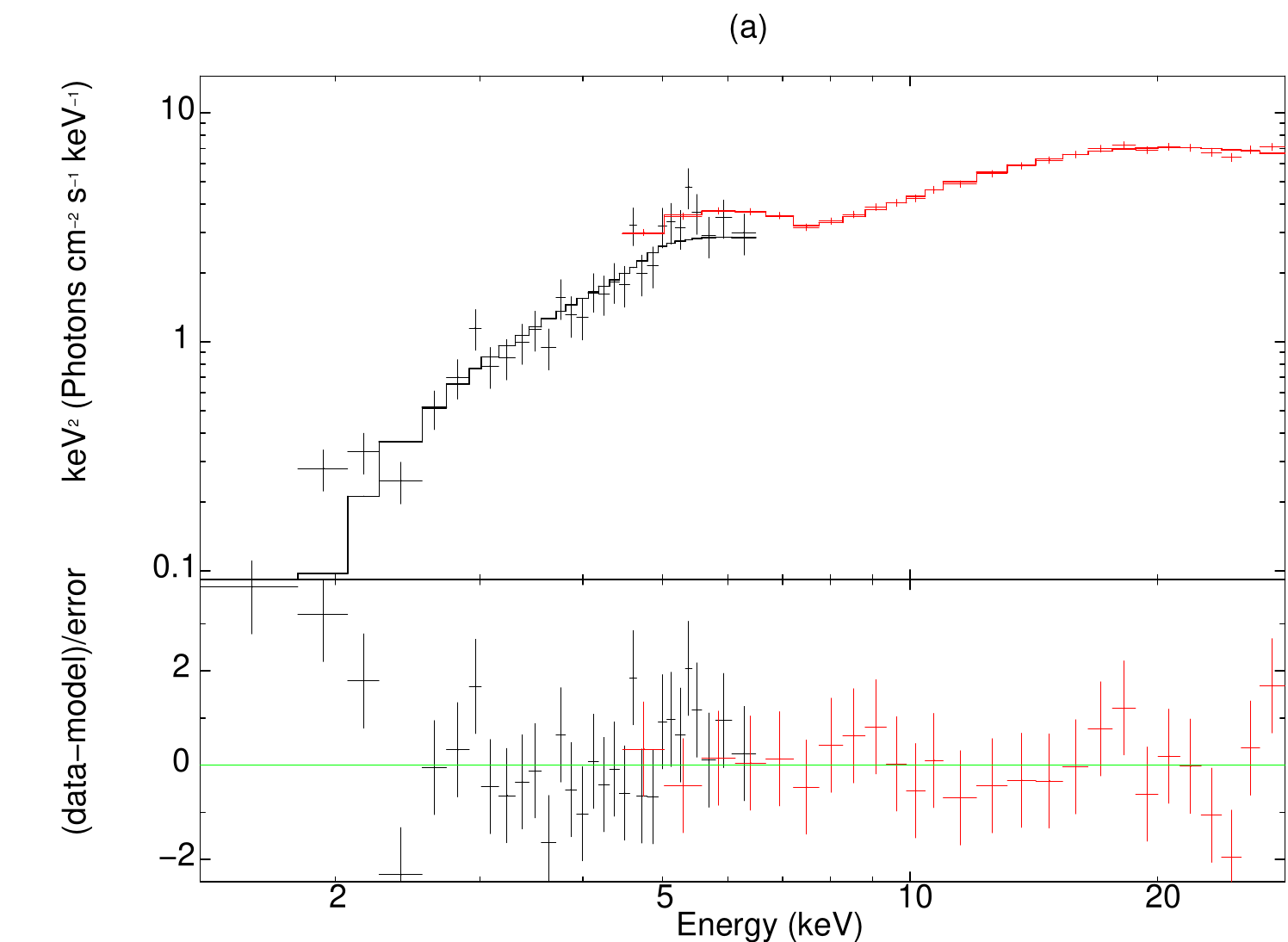}
                     \includegraphics[scale=0.35,angle=-90]{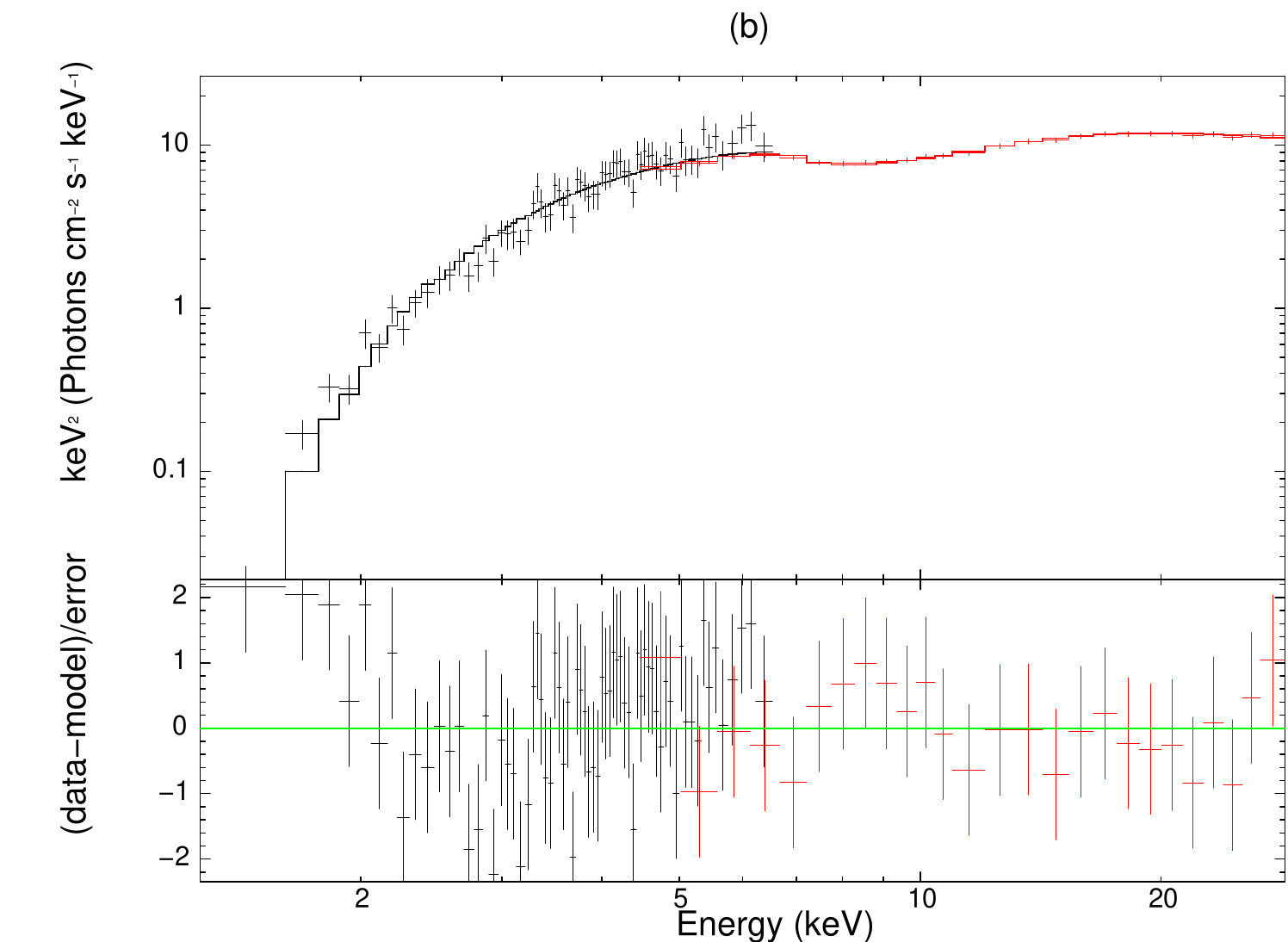}
                     \includegraphics[scale=0.35,angle=-90]{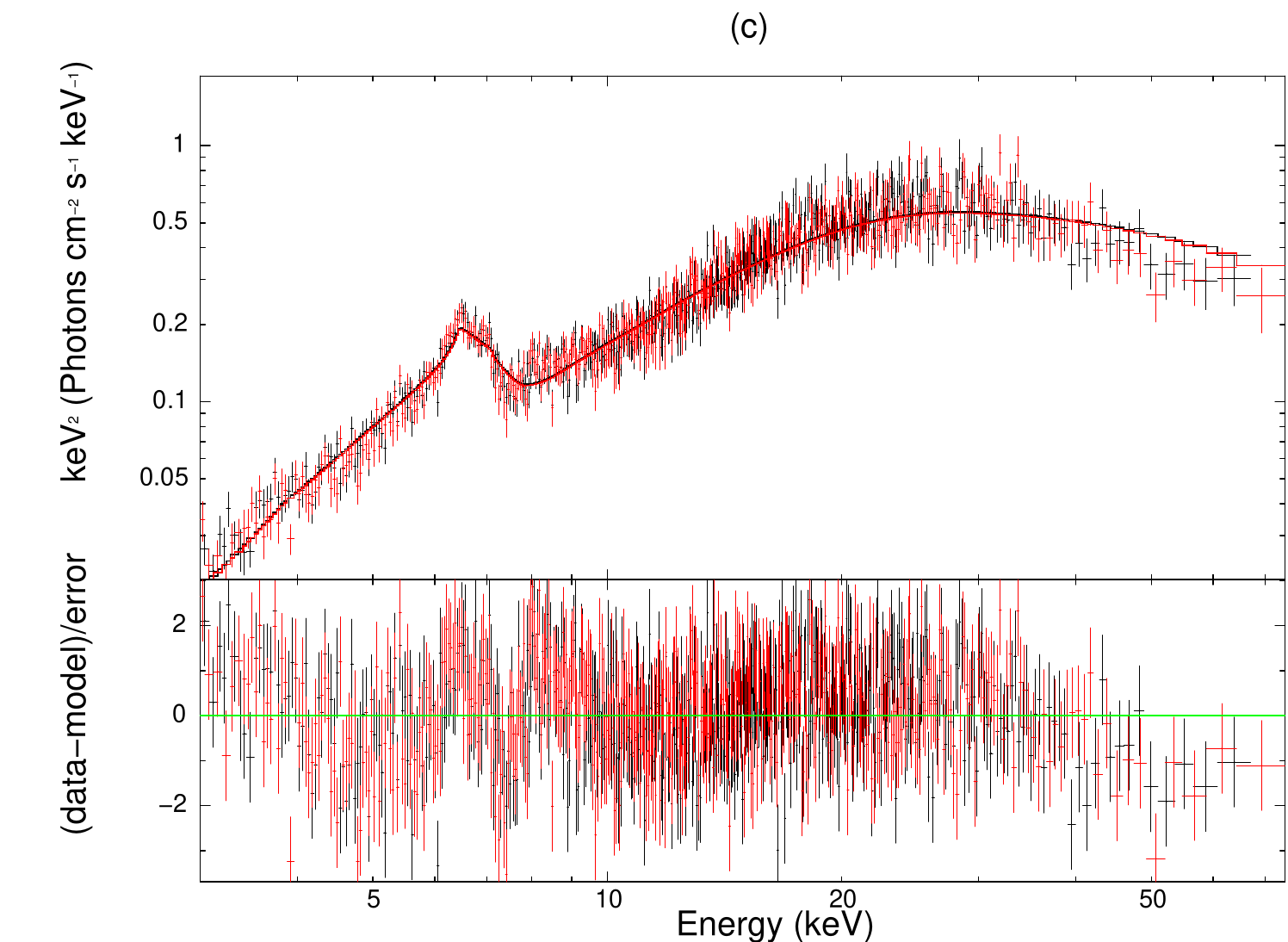}
                     \includegraphics[scale=0.35,angle=-90]{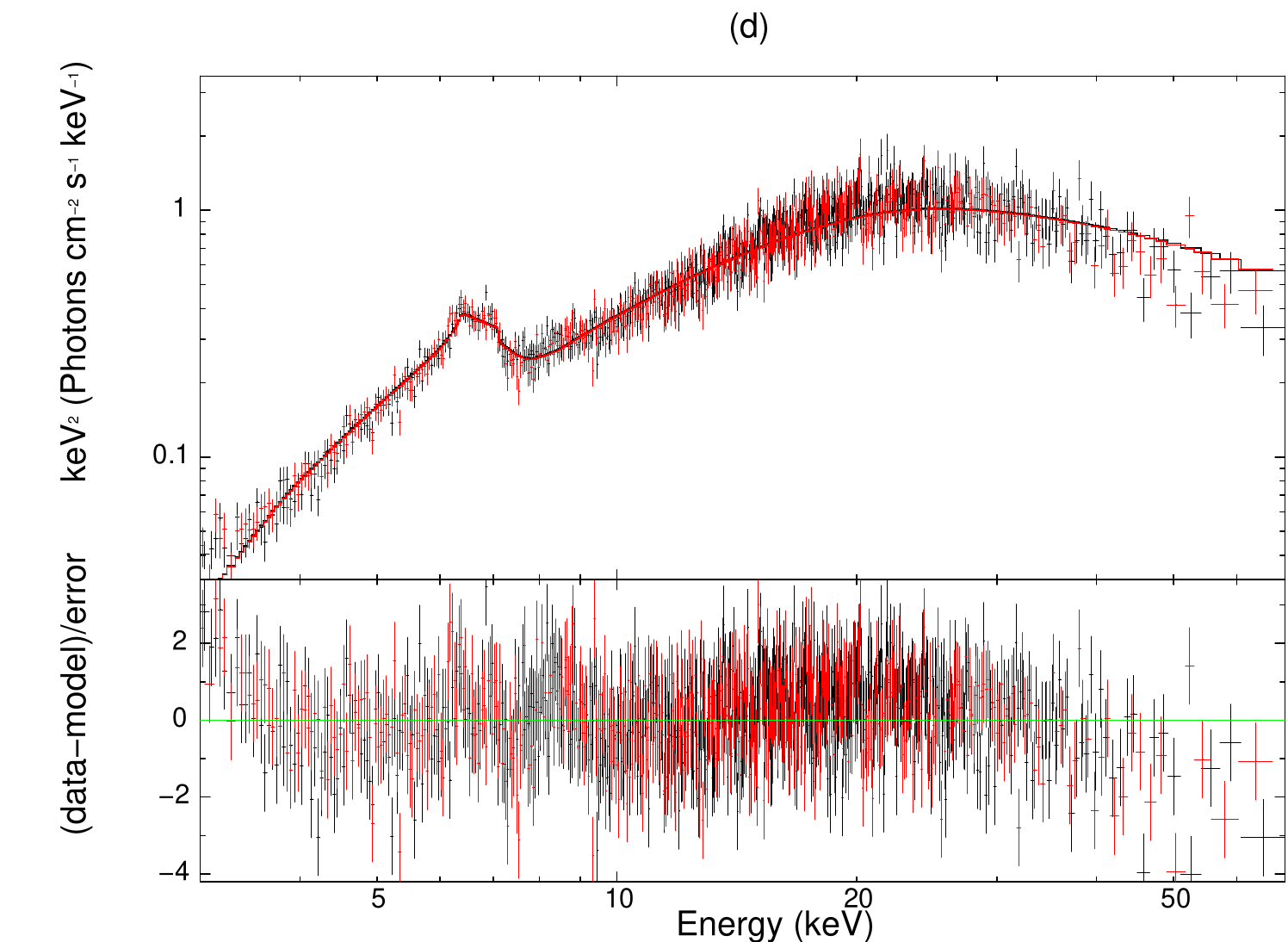}
			\caption{Figures (a) and (b) show the \textit{AstroSat} spectra for segment 8 and segment 10 of flare 2, in medium and high flux states, respectively. These spectra are from the simultaneous fit of 0.7-7 keV SXT and 4-30 keV LAXPC data, modeled using Model A. Figures (c) and (d) display the \textit{NuSTAR} FPMA and FPMB spectra in medium and high flux states, within the energy range of 3--79 keV, also modeled using Model A.}
			\label{Spectra}
		\end{figure*}

 Table~\ref{Table_relxill} summarizes the best-fitting spectral parameters and Figure~\ref{Spectra} shows the fitted spectra for two segments with medium and high flux states respectively for both \textit{AstroSat} and \textit{NuSTAR}. The spectral trends of these parameters for all flares of  \textit{AstroSat}  and  \textit{NuSTAR} with respect to flux are illustrated in Figure~\ref{parameter_flux_relxill}. 

		\begin{figure*}[!htbp]
			\centering
			\includegraphics[scale=0.45]{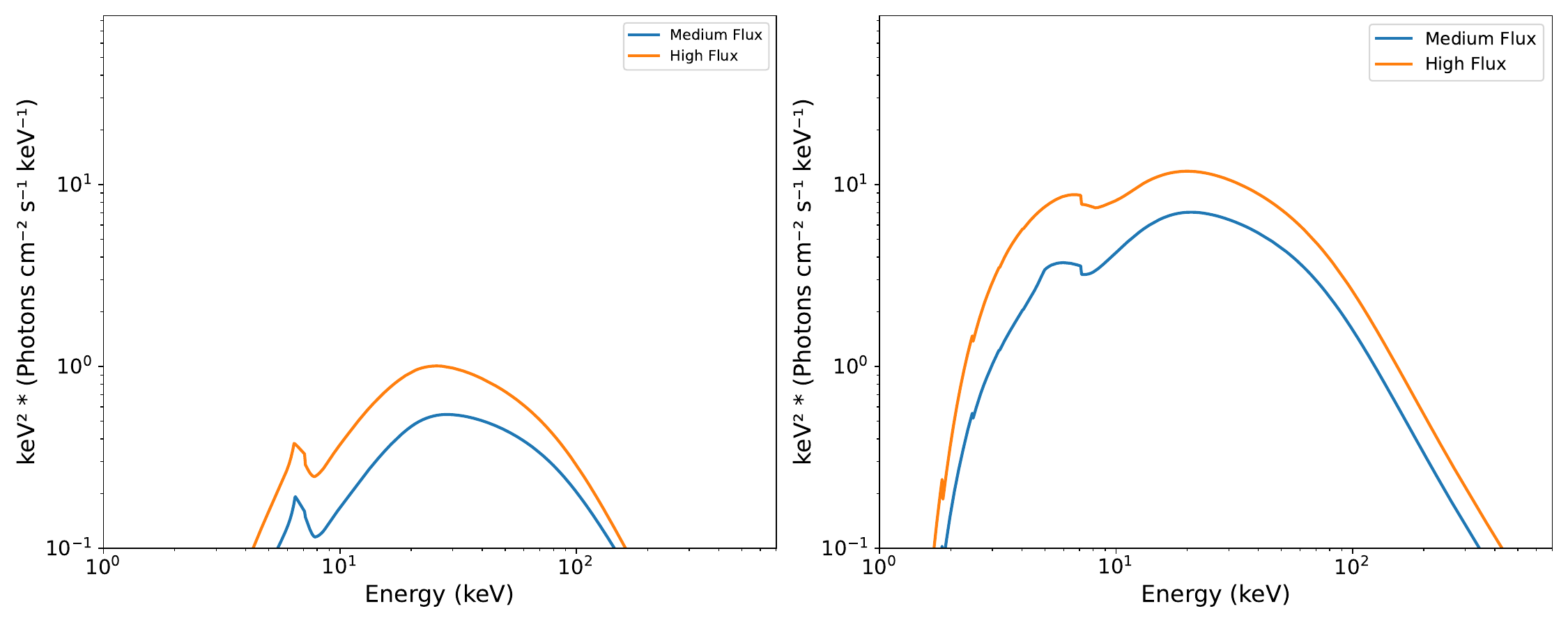} 
			\caption{Model components overlaid using Model A for flare 2  of \textit{AstroSat}  and \textit{NuSTAR} for medium and high flux state. The left panel corresponds to the flux comparison from \textit{NuSTAR}, while the right panel corresponds to the flux comparison from \textit{AstroSat} for medium and high flux states.}
			\label{Fig 15}
		\end{figure*}

\subsection{Spectral Parameter Trends with Flux using Absorption Dominated Model} \label{sec:spectral_absorption}

We began our spectral fitting of all the segments from \textit{AstroSat} and \textit{NuSTAR} by assuming that photons originate from an accretion disc around the black hole and subsequently pass through a Comptonizing medium. The emission from the disc was modeled using the multicolour blackbody model \texttt{diskbb} \citep{Mitsuda1984}, convolved with the Comptonization model \texttt{thcomp} \citep{Zdziarski2020}. To account for interstellar absorption of photons from the source, we included the \texttt{TBabs} model \citep{Wilms2000} in our spectral fitting. Since thcomp is a convolution model, the model energy range was extended from 0.1 to 200 keV using 500 logarithmically spaced bins. In this analysis, we used the spectral index ($\Gamma$) as the fitting parameter in \texttt{thcomp}, which allows either $\Gamma$ or the Thomson optical depth ($\tau$) to be used \citep{Zdziarski1996, Zycki1999, Wilkins2015}. A reflection component \texttt{xillverCp}, was used to model the distant reflection from the disc and and the local obscured absorption features observed in the spectrum was modelled using a photo-ionized absorption model \texttt{zxipcf} \citep{Reeves2008}. The complete XSPEC model expression used was: \texttt{constant}\allowbreak~$\times$\allowbreak~\texttt{TBabs}\allowbreak~$\times$\allowbreak~\texttt{zxipcf}\allowbreak~(%
\texttt{thcomp}\allowbreak~$\times$\allowbreak~\texttt{diskbb}\allowbreak~$+$\allowbreak~\texttt{xillverCp} (hereafter Model B).

The neutral hydrogen column density (\( N_H \)) was fixed at \( 5.0 \times 10^{22} \, \text{cm}^{-2} \), consistent with values reported in previous studies \citep{Zoghbi2017, Koljonen2020}. The covering fraction of the \texttt{zxipcf} component was initially allowed to vary, but as it consistently pegged at unity with an unconstrained upper limit, so we assumed full coverage of the X-ray source by fixing it to 1. The normalization in the \texttt{diskbb} model is given by $\text{diskbb norm} = \left( \frac{R_{\text{in}}}{D_{10}} \right)^2 \cos \theta$, where the distance $D_{10}$ is measured in units of 10 kpcs and $\theta$ is the inclination angle in degrees. We fixed the \texttt{diskbb} normalization to 350, corresponding to an inner disc radius of $1.25\,R_{\mathrm{g}}$,consistent with the disk reaching the ISCO as indicated by previous studies of GRS~1915+105 \citep{Sajad2024}. This also helps stabilize the spectral fitting by avoiding unphysical variations due to degeneracy between disk temperature and normalization, assuming a source distance of 8.6~kpc and an inclination angle of $60^\circ$ \citep{Reid2014}. The \texttt{diskbb} $T_{in}$ was left as a free parameter. The \texttt{xillverCp} was used only as a reflection component, with the reflection fraction fixed at \(-1\). The iron abundance ($A_{\mathrm{Fe}}$), disc density ($\log N$), and inclination angle were fixed to the same values as those used in Section~\ref{sec:spectral_reflection}. 
The photon index of the Comptonization component was allowed to vary freely, while the photon index of the \texttt{xillverCp} component was tied to it. Additionally, the electron temperature of the \texttt{xillverCp} component was tied to that of the Comptonization component and fixed at 400~keV, consistent with the value used by \cite{Sajad2024}, to enable a direct comparison of our results with their analysis.

\begin{figure*}[!htbp]
		
	             \includegraphics[scale=0.35,angle=-90]{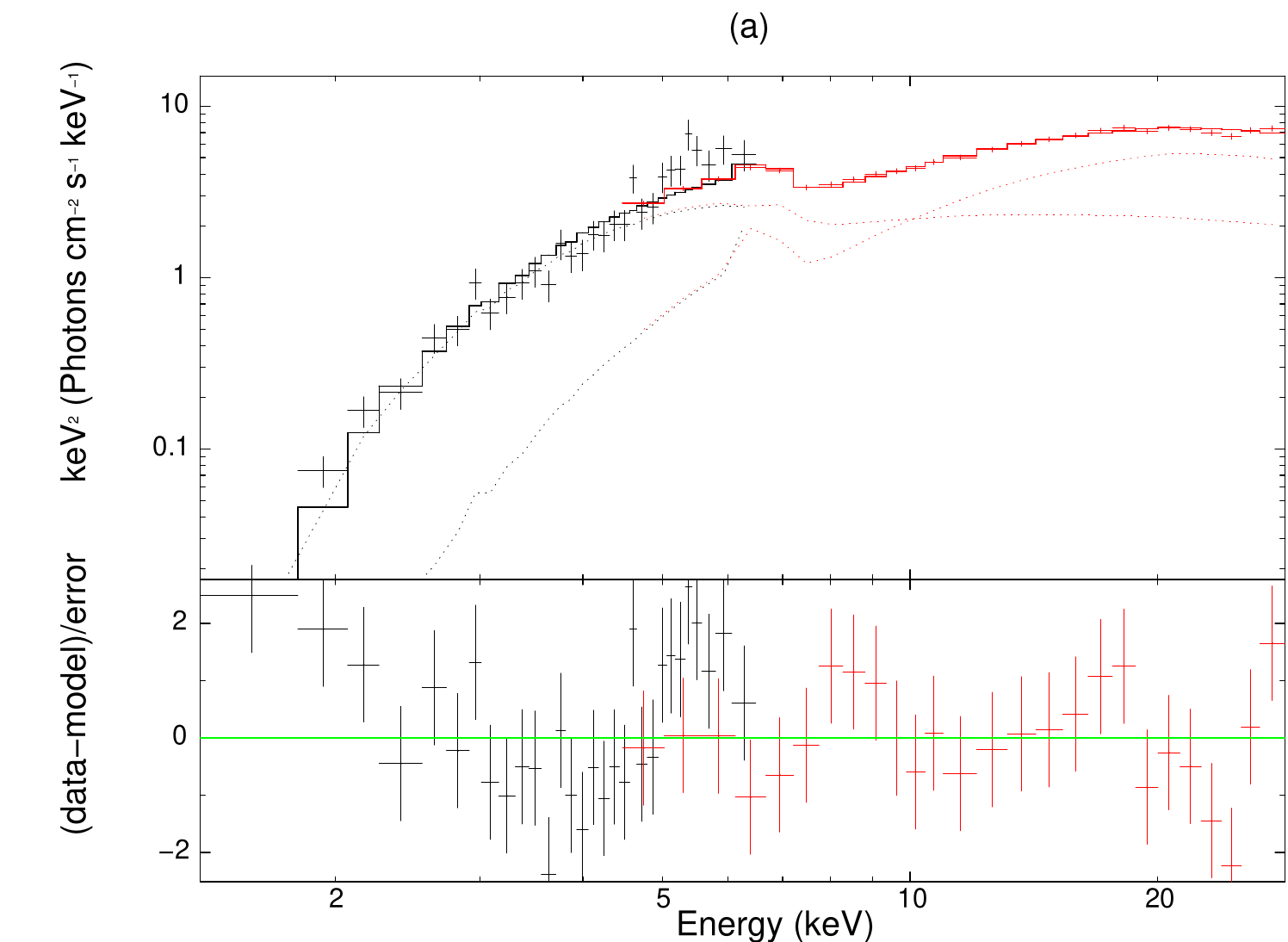}
                     \includegraphics[scale=0.35,angle=-90]{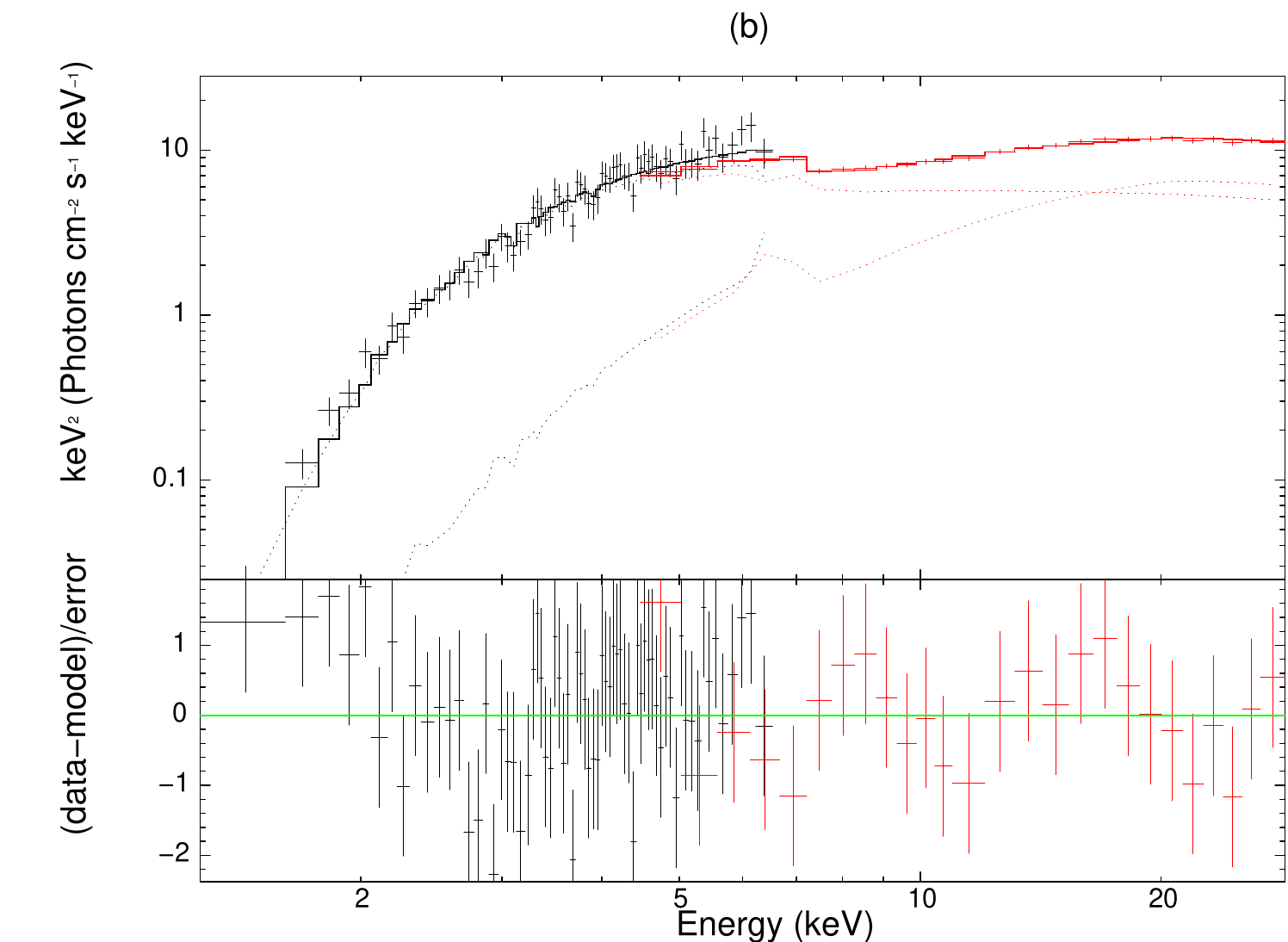}
                     \includegraphics[scale=0.35,angle=-90]{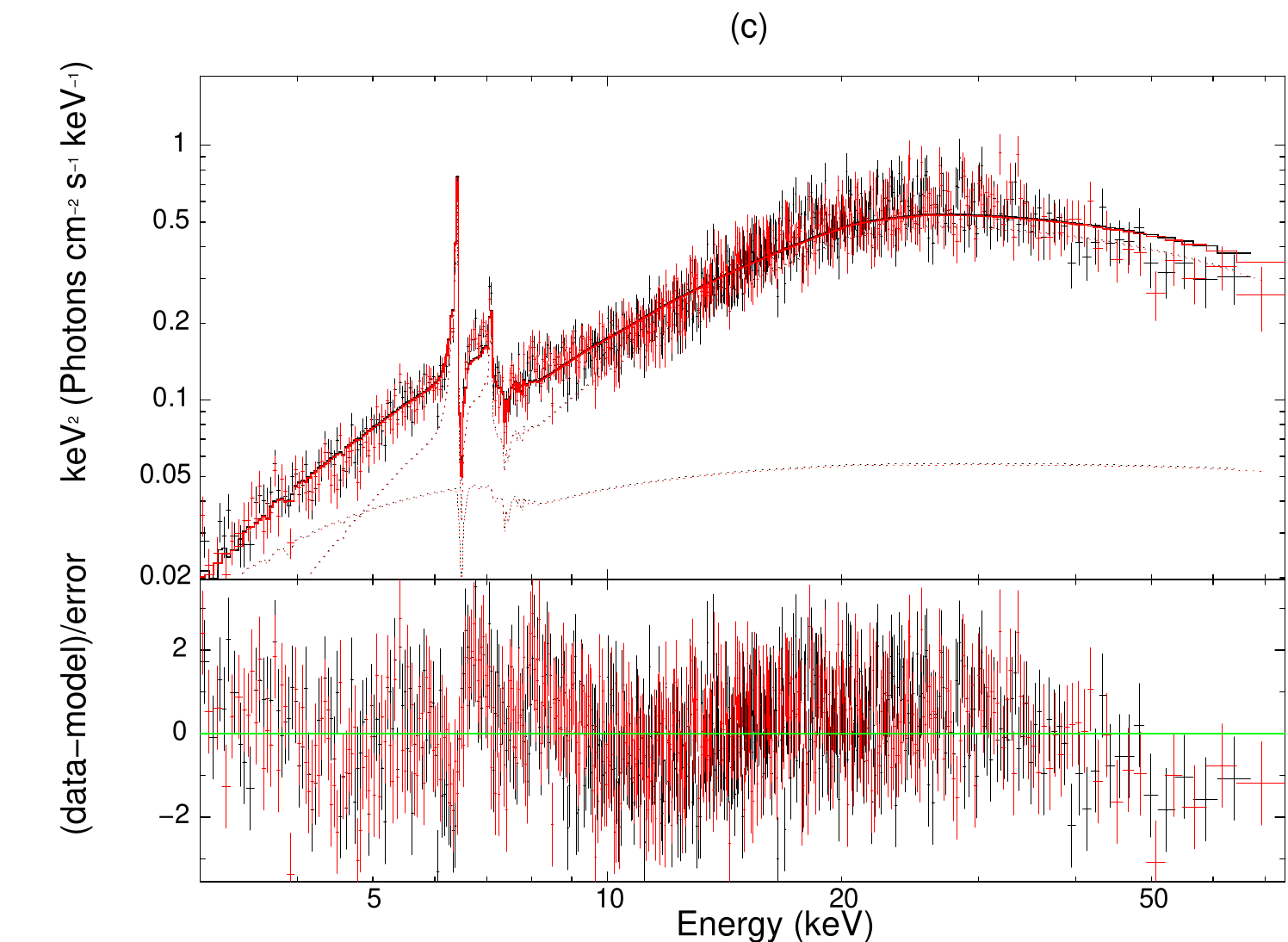}
                     \includegraphics[scale=0.35,angle=-90]{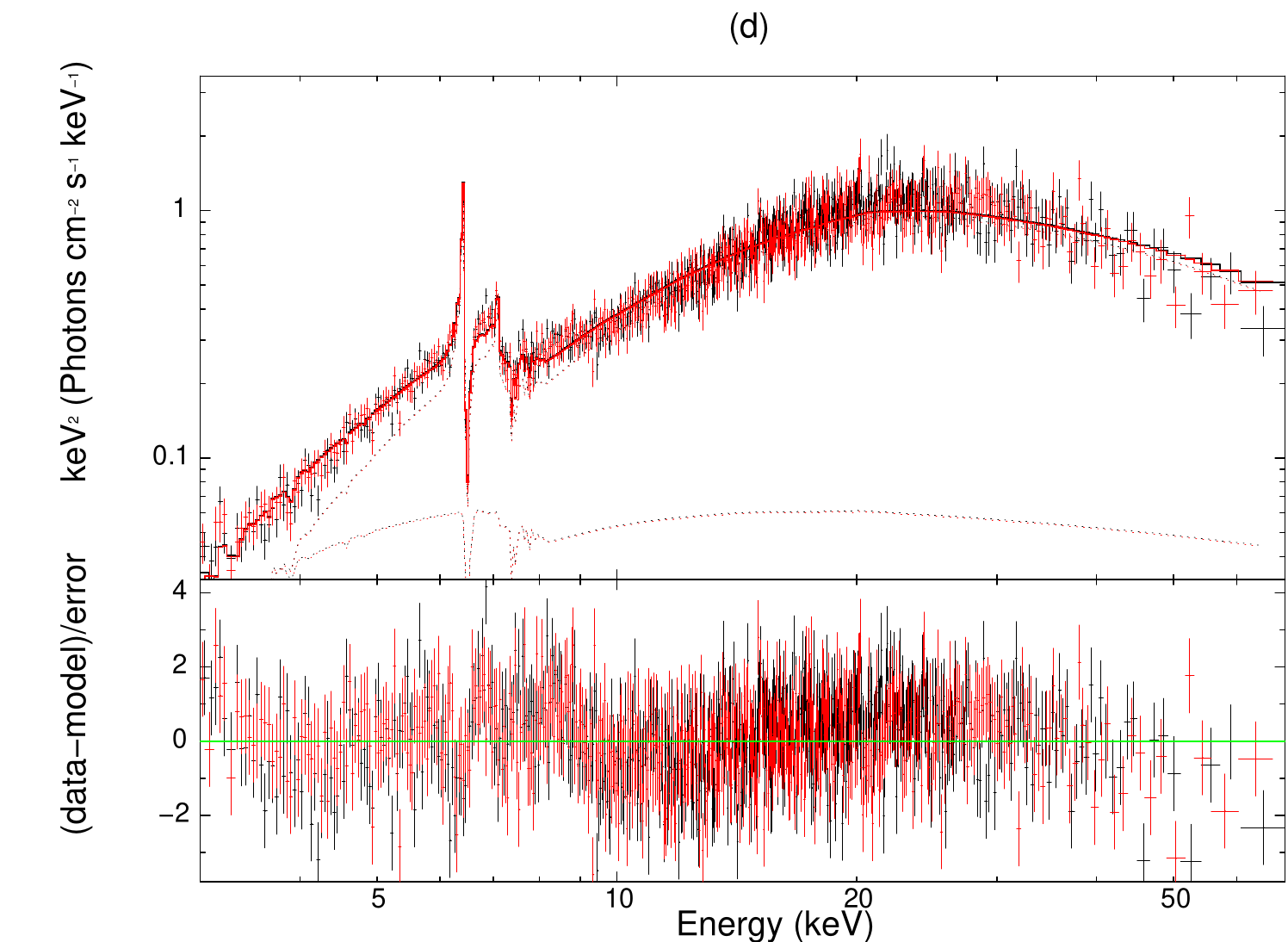}
			
			\caption{Figures (a) and (b) show the \textit{AstroSat} spectra for segment 8 and segment 10 of flare 2, in medium and high flux states, respectively. These spectra are from the simultaneous fit of 0.7--7 keV SXT and 4--30 keV LAXPC data, modeled using Model B. Figures (c) and (d) display the \textit{NuSTAR} FPMA and FPMB spectra in medium and high flux states, within the energy range of 3--79 keV, also modeled using Model B.}
			\label{Spectra_zxipcf}
		\end{figure*}

 We calculated the total unabsorbed flux in the 0.7–30 keV energy range for {\it AstroSat} and 3-79 keV for {\it NuSTAR}, using the \texttt{xspec} model \texttt{cflux}, and for calculating the corresponding luminosities we have assumed a distance of $8.6^{+2.0}_{-1.6}$ kpc. The disc flux was calculated using the \texttt{cflux} applied directly to the \texttt{diskbb} component, as shown in the spectral expression: {constant}$\times${tbabs}$\times${zxipcf}({thcomp}$\times${cflux}$\times${diskbb}~+~{xillverCp}). The fractional disc flux (i.e., disc flux divided by the total unabsorbed flux) is plotted against the total unabsorbed flux, and the disc luminosity is shown as a function of the total unabsorbed luminosity (see Figure~\ref{disk_flux_accretion}).  To examine the relationship between the model parameters and the unabsorbed flux, we used Spearman's  correlation and fitted the data with a power-law of the form:(\( f(x) = A \times x^{-p} \)) where $A$ is the normalization factor and $p$ is the power-law index. The Spearman's correlation coefficients, corresponding p-values, and the power-law fit parameters are presented in Table~\ref{Table 3}. Table~\ref{Table 4} summarizes the best-fitting spectral parameters and Figure~\ref{Spectra_zxipcf} shows the fitted spectra for two segments with medium and high flux states respectively for both \textit{AstroSat} and \textit{NuSTAR}. The spectral trends of these parameters for all flares of  \textit{AstroSat}  and  \textit{NuSTAR} with respect to flux are illustrated in Figure~\ref{parameter_flux_zxipcf}.

		\begin{table*}
			\centering
			\caption{The Spearman's correlation coefficients and p-values, along with the power-law fit parameters (\( f(x) = A \times x^{-p} \)) for the relationship between model parameters and flux using Model B.}
			\begin{tabular}{llcccc}
				\hline
				\textbf{Parameter} & \textbf{A} & \textbf{p} & \textbf{Correlation coefficient} & \textbf{P-value}   \\ 
				\hline
				zxipcf $\log(\xi)$ & $2.91_{-0.11}^{+0.11}$ & $0.02_{-0.02}^{+0.02}$ & -0.35 & $5.07 \times 10^{-4}$  \\  
				thcomp $\Gamma$ &  $1.98_{-0.04}^{+0.04}$ & $-0.04_{-0.01}^{+0.01}$ & 0.47 & $3.89 \times 10^{-2}$  \\  
				zxipcf $N_\mathrm{H}$    & $43.71_{-7.40}^{+7.40}$ & $0.08_{-0.07}^{+0.07}$ & -0.31 & $2.5 \times 10^{-4}$ \\  
				diskbb $T_{\mathrm{in}}$   & $0.40_{-0.02}^{+0.02}$ & $-0.31_{-0.01}^{+0.01}$ & 0.91 & $1.5 \times 10^{-19}$ \\  
				\hline
			\end{tabular}
			
			\label{Table 3}
		\end{table*}

		\begin{figure*}[!htbp]
			\centering
			\includegraphics[scale=0.48]{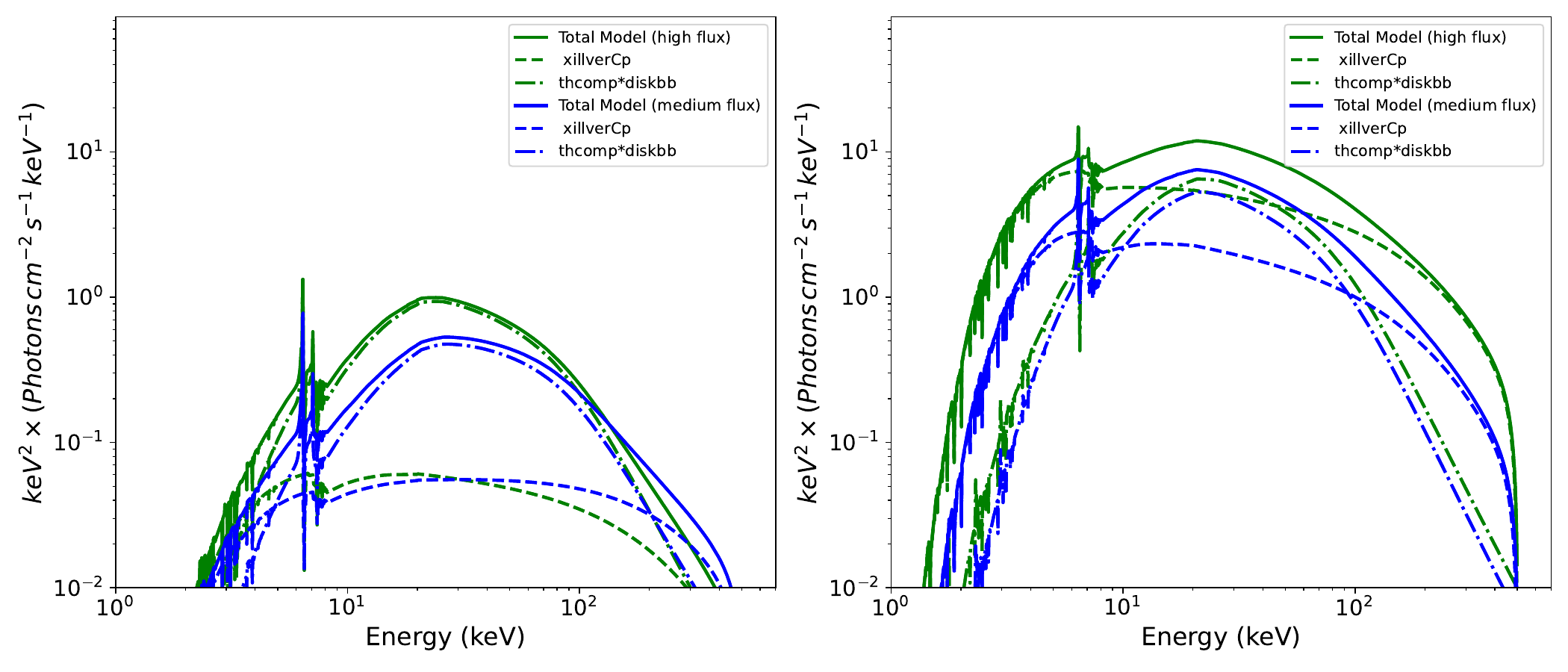} 
			
			\caption{Model components overlaid using Model B for flare 2 of \textit{AstroSat} and \textit{NuSTAR} in medium and high flux states. The left panel corresponds to the flux comparison from \textit{NuSTAR}, while the right panel corresponds to the flux comparison from \textit{AstroSat} for medium and high flux states.}
			
\label{Fig 14}
		\end{figure*}

\section{Results and Discussion} \label{sec:discussion}

\begin{figure}
			\centering
			\includegraphics[width=0.9\columnwidth]{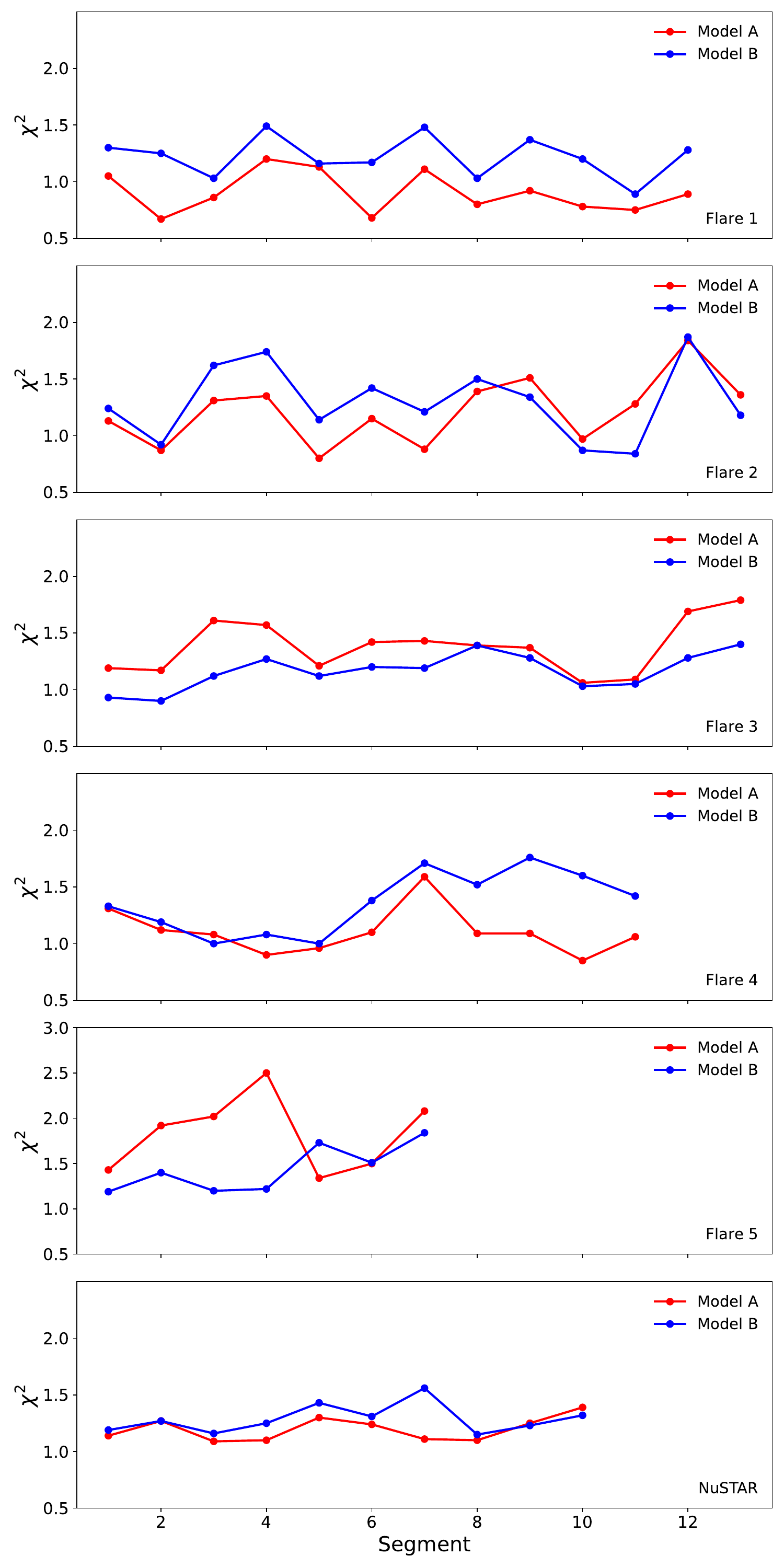}
			\caption{ comparison of $\chi_{reduced}^{2}$ vs segment  for all Flares of \textit{AstroSat} and \textit{NuSTAR} data for reflection dominated model (Model A) and absorption dominated model (Model B)}
			\label{chi_sq}
		\end{figure}

\begin{figure}
	\centering
	\includegraphics[scale=0.35]{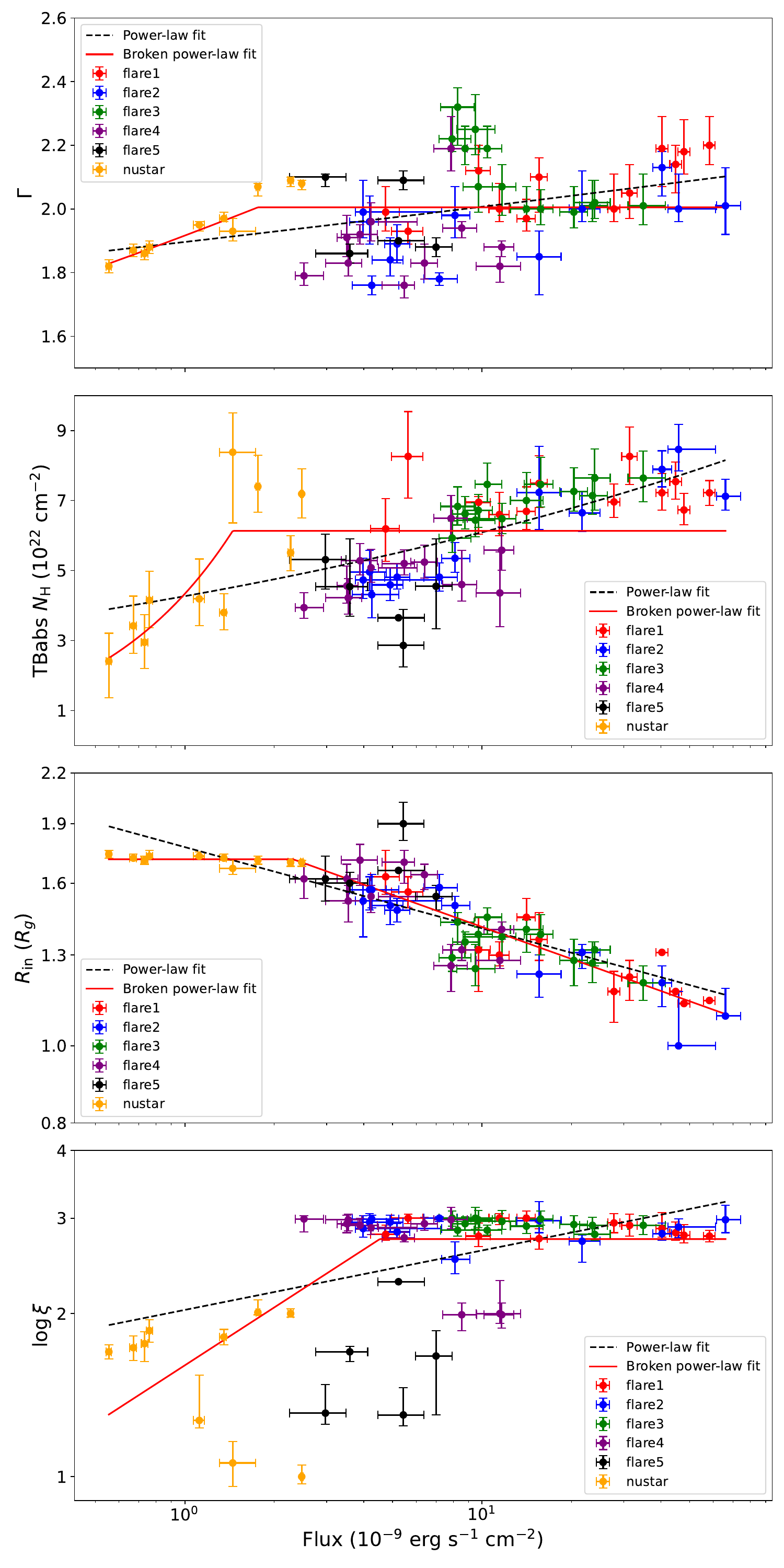}
	
	\caption{	Variation of spectral parameters with unabsorbed flux for all flares, including \textit{NuSTAR} observation using  Model A. The unabsorbed flux is calculated in  0.7--30~keV energy range and reported in units of \(10^{-9}\) erg\,cm\(^{-2}\)\,s\(^{-1}\). From top to bottom, the panels show: (1) \texttt{relxillCp} (\(\Gamma\)) , (2) \texttt{tbabs} column density (\(N_{\mathrm{H}}\)), (3) \texttt{relxillCp}  (\(R_{\mathrm{in}}\)), and (4) \texttt{relxillCp} ionization parameter (\(\log \xi\)).}
	\label{parameter_flux_relxill}
\end{figure}

\begin{figure}
			\centering
			\includegraphics[scale=0.35]{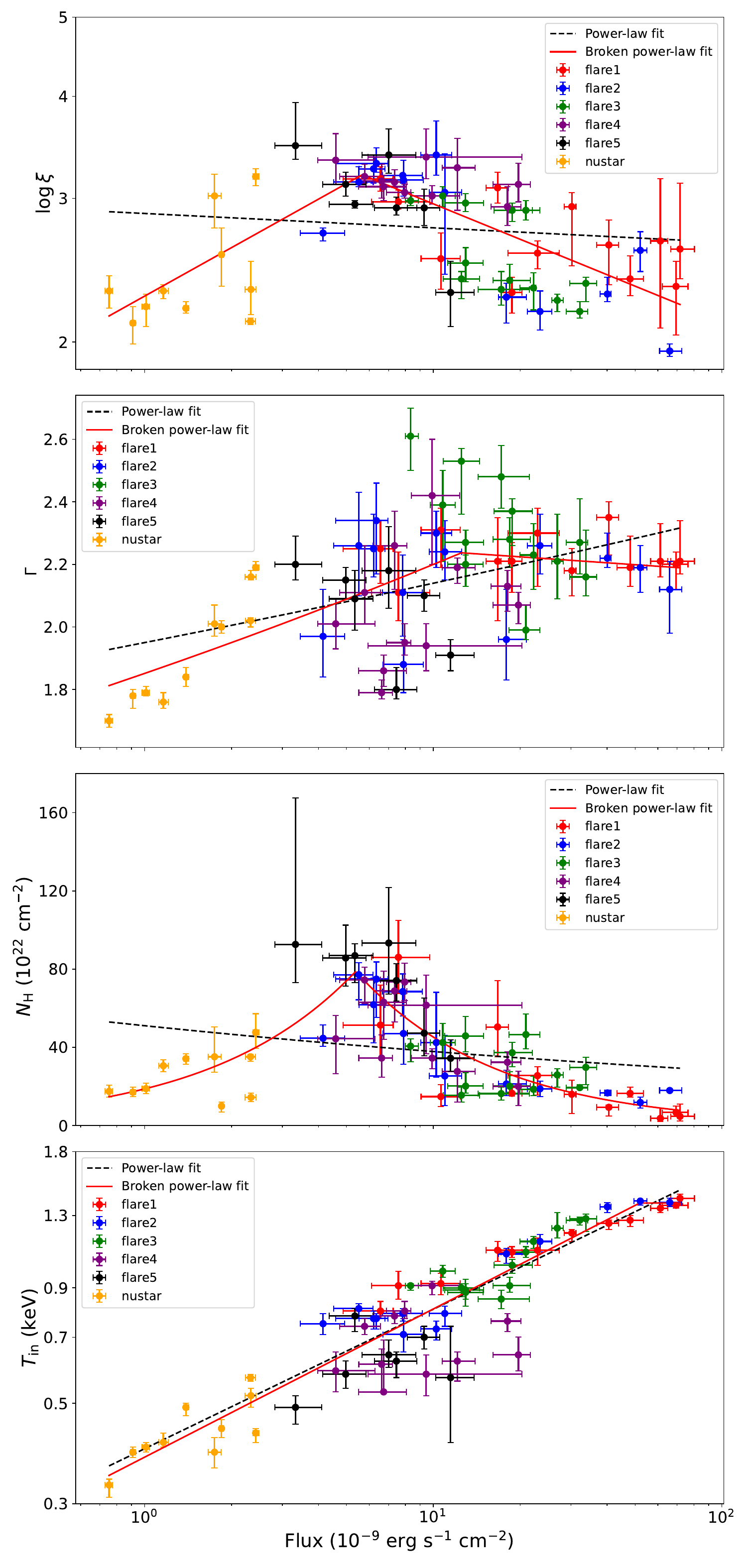}
			\caption{
				Variation of spectral parameters with unabsorbed flux for all flares, including \textit{NuSTAR} observation using  Model B. The unabsorbed flux is calculated in  0.7--30~keV energy range and reported in units of \(10^{-9}\) erg\,cm\(^{-2}\)\,s\(^{-1}\). From top to bottom, the panels show: (1) \texttt{zxipcf} ionization parameter (\(\log \xi\)), (2) \texttt{thcomp} photon index (\(\Gamma\)), (3) \texttt{zxipcf} column density (\(N_{\mathrm{H}}\)), and (4) \texttt{diskbb} inner disk temperature (\(T_{\mathrm{in}}\)).}

			\label{parameter_flux_zxipcf}
		\end{figure}

\cite{Sajad2024} reported the first-time analysis of a single flare from May 2019 using the broadband spectral capabilities of {\it AstroSat} and identified two degenerate spectral models that adequately fit the data. In this work, we extend their findings by incorporating all available {\it AstroSat} observations in which flares were detected during the low anomalous state of the source. The reduced chi-square values obtained from spectral fitting for all reported flares are presented in Figure~\ref{chi_sq}. To validate the spectral models derived from {\it AstroSat} data, we reanalyzed a {\it NuSTAR} spectrum obtained during this dim state and found that it can also be satisfactorily described by the same two models. From our spectral fitting using two different models, we find that for the flaring states observed by \textit{AstroSat}, both models yield statistically comparable fits, at least for the first three flares. However, for flares 4 and 5, deviations from the model fits become more pronounced, with Model A providing a better fit for flare 4, and Model B fitting flare 5 more effectively. In the case of \textit{NuSTAR} data, which do not correspond to a flaring state, both models yield similarly good fits, as seen in Figure~\ref{chi_sq}. Several spectral parameters exhibit a step-like behavior as a function of flux as shown in Figures~\ref{parameter_flux_relxill} and~\ref{parameter_flux_zxipcf}, suggesting that the source undergoes distinct spectral transitions beyond certain flux thresholds (typically between flux values of 5–10  erg\,cm\(^{-2}\)\,s\(^{-1}\)), which vary across parameters. A comparison of fit statistics further reveals that individual flares are better described by different models—for instance, Model A provides a better fit for flare 1, while Model B yields a more satisfactory fit for flare 3. For the remaining flares, the interplay between reflection and absorption appears to influence the spectral shape.  The Spearman's correlation coefficient between the parameters and the corresponding p-values are presented in Table~\ref{Table 3} and  Table~\ref{Table 4}. A correlation coefficient close to 1 indicates a strong correlation, while a coefficient of 0 suggests no correlation. In the interest of space and clarity, we have omitted the correlations between parameters that were not significant.

Notably, in the reflection-dominated model, we observe a positive correlation with a correlation coefficient of 0.40 between the photon index and flux as illustrated in Figure~\ref{parameter_flux_relxill}. This behavior aligns with previous observations of BH-XRBs and AGN (both reflection-dominated and non-reflection-dominated), where the photon index increases with flux \citep{Wang2004,Shemmer2006,Sajad2024}. This positive correlation between the photon index ($\Gamma$) and flux in black hole systems can be attributed to two primary models: the hot accretion flow model and the disk-corona model. In the \textit{hot accretion flow model}, X-rays are produced when soft seed photons interact with the hot accretion flow \citep{zdziarski1998, zdziarski2004, done2007}. These seed photons can originate either from blackbody radiation emitted by the outer accretion disc or from synchrotron radiation generated within the hot flow itself. When the accretion rate increases, the outer disc moves closer to the black hole, leading to greater overlap between the disc and the hot flow. This increased overlap boosts the flux of seed photons reaching the hot flow, intensifying its cooling. The enhanced cooling steepens the X-ray spectrum, resulting in a positive correlation between the photon index ($\Gamma$) and flux \citep{zdziarski1999}. The \textit{disc-corona model} presents an alternative scenario, where X-rays are emitted by a hot corona located above a thin accretion disc that extends all the way to the innermost stable circular orbit (ISCO) \citep{liang1977, galeev1979, haardt1991, stern1995, merloni2001,  schnittman2013}. As the accretion rate increases, a larger fraction of the energy is dissipated within the disc rather than in the corona. This redistribution leads to a reduction in the energy of hard X-rays emitted by the corona, softening the X-ray spectrum and naturally producing a positive index-flux correlation \citep{cao2009}. This results in higher coronal temperatures and lower photon indices. Therefore, the spectral variability in these systems is primarily driven by changes in the spectral slope of the primary X-ray continuum.

The launching mechanisms of accretion disc winds remain a topic of debate, with three main driving processes proposed: thermal, radiation-pressure, and magnetic forces. Evidence suggests that all three mechanisms may play a role, and in some cases, hybrid winds involving both thermal and magneto-hydrodynamic (MHD) components have been observed \citep{Neilsen2012}. In source like GRO J1655-40, studies have shown that a single absorption component is insufficient to explain the wind, with multiple absorption components being required to model the observed spectral features \citep{Miller2006a,Miller2015,Kallman2009}. Some findings suggest that thermal winds arise from larger radii, while MHD-driven winds originate from innermost parts of the disc. Our results from both the reflection-dominated and absorption-dominated modeling show a difference in the ionization parameter for both the ionized absorption and reflection components, as well as across lower and higher flux levels. As evident from the Figure~\ref{parameter_flux_relxill} and~\ref{parameter_flux_zxipcf} , we observe a cutoff that exhibits a step-function-like behavior in each of the model parameter correlations using broken power law. This suggests that a single-component wind is insufficient to explain the observed features. Instead, we argue that multiple ionized winds are required, where the inner regions may correspond to reflection-dominated winds, while the outer regions align with an absorption-dominated wind. This supports the idea that winds in accretion systems are complex, structured, and not purely thermal or magnetic in origin. To quantitatively assess the statistical significance of the step-function-like cutoffs observed in the correlations of spectral parameters, we performed F-tests comparing single power-law (PL) and broken power-law (BPL) fits. The results, summarized in Tables~\ref{tab:f_test_model1} and \ref{tab:f_test_model2}, show that for Model A, the F-test for the inner disk radius $R_{\rm in}$ and ionization parameter $\log \xi$ yields large positive F-values (8.28 and 5.70) with very small p-values (0.0007 and 0.0055), confirming the presence of step-like trends. However, for $\Gamma$ and $N_{\rm H}$ from Model A, negative F-values and high p-values (1.0) indicate that the BPL does not significantly improve the fit, and a single PL is sufficient. In the case of Model B, the parameters $N_{\rm H}$, $\log \xi$, and $\Gamma$ show very small p-values ($p \ll 0.04$), suggesting a statistically significant improvement of the BPL fits over the PL fits. In contrast,for parameters like   $T_{\rm in}$ from Model B, the F-test does not show a significant improvement ($p \sim 0.6$), consistent with the step-like behavior being weak or absent. These results confirm that the observed cutoffs in some spectral parameters versus flux are statistically robust, supporting the use of broken power-law modeling.

In the reflection-dominated scenario, Our analysis revealed that the inner disk radius is consistently very close to the ISCO and shows a negative correlation with the 0.7–30 keV unabsorbed flux as evident from Figure~\ref{parameter_flux_relxill}. The steeper emissivity index (see Table \ref{Table_relxill}) observed in all the segments further supports the notion that the disk is primarily illuminated in its inner regions. Furthermore, the rise in the column density of the line of sight absorber may also be linked to the disk radius approaching the ISCO. As the inner disk edge moves closer to the ISCO, the line of sight will pass through denser regions of the accretion disk, which could lead to increased absorption, thereby contributing to the observed rise in the column densities. Also as the inner disk edge moves closer to the ISCO, more and more high energy photons from the X-ray emitting region interact with the disc and its atmosphere, and are reflected either from the disc or the local obscuring medium thus enhancing the reflection fraction. Thus the enhanced flux observed in the sharp flares  may result from the X-ray emitting source moving closer and farther from the central BH, since there is a strong indication of centrally concentrated primary X-ray source from our observations of steep emissivity profile and broad iron line. This kind of variation can be explained on the basis of relativistic effects such as light-bending model proposed by \citet{Miniutti2004}. Thus the observed flares may be as a consequence of interaction of disc and the X-ray emitting source.

\begin{figure*}
			\centering
			\includegraphics[width=2.2\columnwidth]{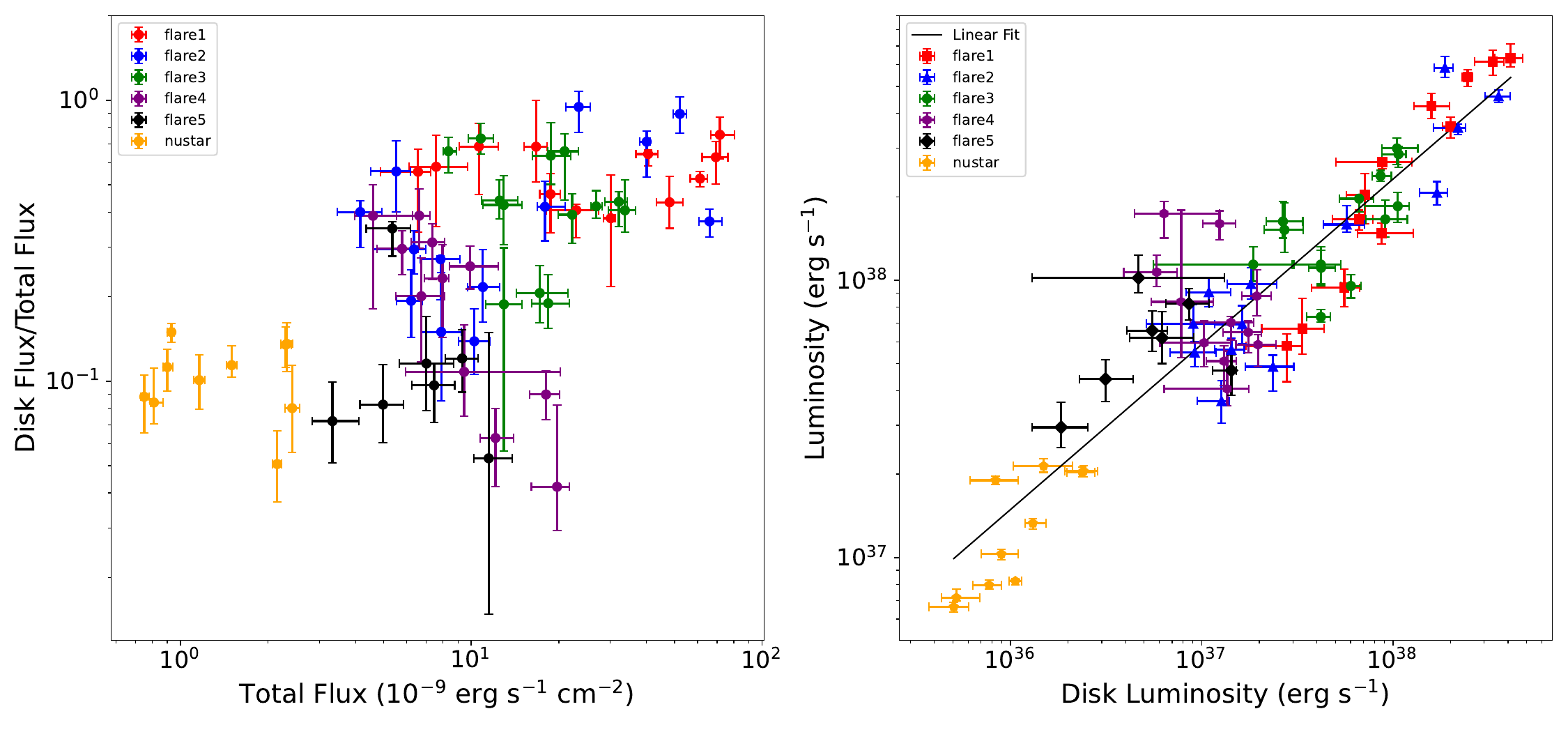} 
			\caption{Right panel:  Total unabsorbed Luminosity   is plotted as a function of  Disk Luminosity for all flares of \textit{AstroSat}  and \textit{NuStar} data and  in left panel fractional disk flux is plotted with total  unabsorbed flux  for all flares of \textit{AstroSat} data with \textit{NuStar} data  for Model B.}
			\label{disk_flux_accretion}
		\end{figure*}

As shown in the left panel of Figure ~\ref{disk_flux_accretion}, the disk flux remains consistently proportional to the total flux up to a certain flux level. Beyond this point, the disk flux increases, indicating that the disk becomes the primary contributor to the total flux once it surpasses a certain threshold. Therefore, we can conclude that the disk is the main contributor to the observed flares. Also from right panel of the figure the disc luminosity contributes to the total luminosity, and we found a strong correlation (spearman rank correlation coefficient of 0.89) between the two thus indicating the strong role of disc in the observed flares. Additionally, we observed a negative correlation between the column density of the ionized absorber \textit{zxipcf} \(N_\mathrm{H}\) and flux. This suggests that flares are not only attributed to enhanced emission in the disk but are also influenced by changes in the local absorption properties of the system. Hence the flares seen in the light curves are not merely due to a change in the absorption column density but are to be driven by combination of increased disk flux, elevated accretion rate, and reduced absorption.

To confirm the results from the \textit{AstroSat} analysis, we re-analyzed a \textit{NuSTAR} spectrum from the same dim state and confirmed that it can also be modeled using the same approach. In a previous study, \citet{Koljonen2020} examined the time averaged spectra of the \textit{NuSTAR} observation used in this study with a model combination {\sc phabs × pcfabs × (xillverCp + xillverCp)}. We have performed time-resolved spectroscopy to observe the temporal variation of spectral parameters of the same {\it NuSTAR} observation and modelled it using both reflection dominated and ionized absorption dominated models. Their time averaged spectra revealed harder power law index of 1.6 whereas in our modelling it varied between 1.7 to 2.2 in all the segments for both of our models. Their neutral partially absorbed covering fraction was 0.6 while we have assumed the full covering of our ionized partial covering absorber. Further the two ionization zones from their analysis with different ionization by using two {\it xillverCp} components are also verified from our ionized absorption model with different ionization for the {\it zxipcf} and {\it xillverCp} models. 
			
In  reflection dominated model, the inner radius of the disc appears to move closer to the black hole, enhancing relativistic light-bending effects and increasing the reflection fraction. Since the source remains reflection-dominated across all flares despite variations in intrinsic flux, this implies that the flares are intrinsic in nature. These results are analogous to the reflection-dominated low-flux spectra observed in AGNs such as MCG–6-30-15 \citep{Fabian2003}, 1H0707–495 \citep{Fabian2004}, and 1H0419–577 \citep{Fabian2005}, where strong light bending near the central black hole enhances reflection. In contrast, Model B suggests that changes in the accretion rate and an increased contribution of the disc to the total flux (as seen in Figure~\ref{disk_flux_accretion}) are responsible for the flaring behavior, with no significant change in the absorber properties, which remain consistent throughout. The decrease in local absorption with increase in flux may be attributed to stronger radiation pressure or enhanced magnetic activity at higher accretion rates, which can displace the ionized material near the corona. This reduces the absorbing column density in the vicinity of the source, allowing more intrinsic emission to escape and leading to an increase in the observed flux. So we can conclude that the flux variation is due to intrinsic variation of the source and cannot be attributed only to variations in the absorber. These findings suggest that the flare activity during the anomalous low-flux state of GRS 1915+105  may be  governed by a dynamic interplay between enhanced intrinsic emission and a variable, likely stratified absorber near the inner accretion flow.

Finally, we note that the characteristic timescale of the \textit{AstroSat} flares ($\sim 200$--$500$~s) is significantly shorter than that of the \textit{NuSTAR} flare ($\sim 30,000$~s), with \textit{AstroSat} flares exhibiting higher intensity, whereas the \textit{NuSTAR} flare is associated with comparatively lower flux. Although these variabilities (in timescales) are different, our time-resolved spectroscopy analysis indicates that the spectral parameters derived from both types of flares remain broadly consistent, arguing for a similar underlying emission process. Interestingly, flares with similarly long timescales ($\sim 20,000$~s) have been reported in Insight–HXMT observations of GRS 1915+105 during June 2019 \citep{Kong2021}. \citet{Kong2021} interpreted these flares in terms of variable absorption and jet--wind transitions regulated by large-scale magnetic fields. They also suggested different behaviour of the flare spectrum during bright and faint epochs. This reveals the presence of two distinct characteristic timescales: a long-term timescale reflecting the gradual decay of intense flare events, and a short-term timescale corresponding to individual epochs (peaks) in the broad flare. Notably, this rapid, short-timescale variability is predominantly observed when the source flux exceeds a certain threshold, and is notably absent below this level, as evidenced by step-like behaviour in flux--parameter correlations. 
Despite these observed differences in flux and timing behaviour, the spectral characteristics of \textit{AstroSat} and \textit{NuSTAR} flares display broad consistency, indicating a shared underlying physical mechanism modulated by varying accretion states. These findings suggest that multiple mechanisms contribute to the complex flare processes observed in GRS~1915+105.

		\section{ACKNOWLEDGEMENTS}
		This study utilizes data from the \textit{AstroSat} mission, conducted by the Indian Space Research Organisation (ISRO), and archived at the Indian Space Science Data Centre (ISSDC). Specifically, data from the LAXPC and SXT instruments onboard \textit{AstroSat} were analyzed. We extend our gratitude to the LAXPC and SXT Payload Operation Center (POC) teams at TIFR Mumbai for providing the essential software tools for this analysis. SA, BM, and SB extend their gratitude to IUCAA, Pune, for providing support and facilitating regular visits under its visitor program, during which a portion of this work was completed. V.J. acknowledges the support provided by the Department of Science and Technology (DST) under the ‘Fund for Improvement of S \& T Infrastructure (FIST)’ program (SR/FST/PS-I/2022/208). V.J. also thanks IUCAA, Pune, India, for the Visiting Associateship.Additionally, this research has benefited from the use of data and software provided by the High Energy Astrophysics Science Archive Research Center (HEASARC), a service of the Astrophysics Science Division at NASA/GSFC.
		
		\section{Data availability}
	     The data used in this study can be accessed from the \textit{AstroSat}-ISSDC archive at 
		\url{http://astrobrowse.issdc.gov.in/astro_archive/archive/Home.jsp} and the\textit{NuSTAR} archive at 
		\url{https://heasarc.gsfc.nasa.gov/docs/nustar/nustar_archive.html}. The software employed for data 
		analysis is available for download from the HEASARC website at 
		\url{https://heasarc.gsfc.nasa.gov/lheasoft/download.html}.
		
		\bibliography{references}{}       
        \bibliographystyle{aasjournalv7}

		\appendix
		
		\setlength{\tabcolsep}{4pt} 
		\begin{deluxetable*}{cccccccccc}
		    \digitalasset
			\tablenum{4} 
			
			\tablecaption{Best-fit spectral parameters for the Model A  applied to AstroSat data and NuSTAR data in the 0.3–30.0 keV energy range.The flux reported here is the unabsorbed flux in the energy range 0.7--30 keV  measured in $10^{-9}$ erg cm$^{-2}$ s$^{-1}$.}
			\label{Table_relxill}
			\tablewidth{0pt}
			\tabletypesize{\scriptsize} 
			\tablehead{
				\colhead{\small \textbf{Seg}} &
				\colhead{\small \textbf{TBabs}} &
				\colhead{\small \textbf{relxillCp}} &
				\colhead{\small \textbf{relxillCp}} &
				\colhead{\small \textbf{relxillCp}} &
				\colhead{\small \textbf{relxillCp}} &
				\colhead{\small \textbf{relxillCp}} &
				\colhead{\small \textbf{constant}} &
				\colhead{\small $\boldsymbol{\chi_{red}^{2}}$} &
				\colhead{\small \textbf{Flux}} \\
				\colhead{} &
				\colhead{\small $N_{\rm H}$} &
				\colhead{ \small $R_{\text{in}}$ (ISCO)} &
				\colhead{\small Index} &
				\colhead{\small $\Gamma$} &
				\colhead{\small $\log \xi$} &
				\colhead{\small norm} &
				\colhead{\small factor} &
				\colhead{} &
				\colhead{\small $(10^{-9} \,\text{erg/cm}^2/\text{s})$}
			}
			\startdata
			\multicolumn{10}{c}{\texttt{Flare 2}} \\
			1 & $4.96_{-0.53}^{+0.62}$ & $1.57_{-0.07}^{+0.09}$ & $>7.54$ & $1.96_{-0.06}^{+0.08}$ & $2.95_{-0.11}^{+0.12}$ & $0.03_{-0.01}^{+0.001}$ & $0.99_{-0.12}^{+0.14}$ & 1.13 & $4.24_{-0.07}^{+0.07}$ \\
			2 & $4.73_{-0.52}^{+0.63}$ & $1.52_{-0.08}^{+0.15}$ & $>6.55$ & $1.99_{-0.08}^{+0.10}$ & $2.87_{-0.10}^{+0.15}$ & $0.03_{-0.01}^{+0.01}$ & $1.20_{-0.15}^{+0.19}$ & 0.87 & $3.96_{-0.06}^{+0.06}$ \\
			3 & $4.31_{-0.66}^{+0.72}$ & $1.58_{-0.07}^{+0.09}$ & $>7.77$ & $1.76_{-0.03}^{+0.03}$ & $2.99_{-0.05}^{+0.05}$ & $0.03_{-0.01}^{+0.001}$ & $0.97_{-0.13}^{+0.19}$ & 1.32 & $4.25_{-0.07}^{+0.07}$ \\
			4 & $4.81_{-0.35}^{+0.39}$ & $1.48_{-0.04}^{+0.05}$ & $>8.50$ & $1.89_{-0.06}^{+0.05}$ & $2.83_{-0.08}^{+0.10}$ & $0.04_{-0.01}^{+0.001}$ & $1.12_{-0.09}^{+0.10}$ & 1.36 & $5.19_{-0.06}^{+0.06}$ \\
			5 & $4.59_{-0.46}^{+0.49}$ & $1.50_{-0.05}^{+0.09}$ & $>7.94$ & $1.84_{-0.05}^{+0.07}$ & $2.95_{-0.12}^{+0.09}$ & $0.03_{-0.00}^{+0.01}$ & $1.05_{-0.11}^{+0.10}$ & 0.80 & $6.23_{-0.08}^{+0.08}$ \\
			6 & $5.35_{-0.40}^{+0.45}$ & $1.50_{-0.04}^{+0.08}$ & $>7.83$ & $1.98_{-0.07}^{+0.09}$ & $2.52_{-0.14}^{+0.20}$ & $0.08_{-0.01}^{+0.02}$ & $1.09_{-0.10}^{+0.11}$ & 1.15 & $8.11_{-0.09}^{+0.09}$ \\
			7 & $3.45_{-0.67}^{+1.43}$ & $1.42_{-0.09}^{+0.08}$ & $7.67_{-1.20}^{+2.08}$ & $2.20_{-0.02}^{+0.03}$ & $1.16_{-0.18}^{+0.53}$ & $0.24_{-0.03}^{+0.04}$ & $1.10_{-0.11}^{+0.13}$ & 1.10 & $10.68_{-0.12}^{+0.13}$ \\
			8 & $6.65_{-0.54}^{+0.62}$ & $1.31_{-0.04}^{+0.06}$ & $>8.42$ & $2.00_{-0.04}^{+0.12}$ & $2.72_{-0.23}^{+0.07}$ & $0.19_{-0.02}^{+0.08}$ & $1.30_{-0.11}^{+0.13}$ & 1.39 & $21.73_{-0.24}^{+0.24}$ \\
			9 & $7.89_{-0.50}^{+0.53}$ & $1.20_{-0.07}^{+0.07}$ & $7.29_{-0.74}^{+0.89}$ & $2.13_{-0.10}^{+0.05}$ & $2.81_{-0.08}^{+0.13}$ & $0.42_{-0.09}^{+0.08}$ & $1.00_{-0.07}^{+0.07}$ & 1.00 & $40.54_{-0.43}^{+0.43}$ \\
			10 & $7.12_{-0.40}^{+0.49}$ & $>1.17$ & $7.12_{-0.59}^{+0.72}$ & $2.01_{-0.09}^{+0.12}$ & $2.98_{-0.16}^{+0.19}$ & $0.47_{-0.07}^{+0.14}$ & $0.97_{-0.06}^{+0.07}$ & 0.97 & $65.93_{-0.69}^{+0.70}$ \\
			11 & $8.46_{-0.61}^{+0.72}$ & $>1.15$ & $6.22_{-0.73}^{+0.64}$ & $2.00_{-0.04}^{+0.11}$ & $2.89_{-0.13}^{+0.10}$ & $0.39_{-0.03}^{+0.12}$ & $0.97_{-0.08}^{+0.07}$ & 0.97 & $45.84_{-0.49}^{+0.50}$ \\
			12 & $7.23_{-1.06}^{+1.32}$ & $1.23_{-0.08}^{+0.08}$ & $7.96_{-0.98}^{+1.44}$ & $1.85_{-0.12}^{+0.08}$ & $2.66_{-0.15}^{+0.20}$ & $0.24_{-0.02}^{+0.07}$ & $1.12_{-0.11}^{+0.10}$ & 0.93 & $51.33_{-0.54}^{+0.55}$ \\
			13 & $4.81_{-0.40}^{+0.41}$ & $1.58_{-0.06}^{+0.05}$ & $>8.80$ & $1.78_{-0.01}^{+0.03}$ & $3.00_{-0.04}^{+0.02}$ & $0.04_{-0.00}^{+0.01}$ & $1.01_{-0.09}^{+0.10}$ & 1.36 & $7.19_{-0.09}^{+0.09}$ \\
			\multicolumn{10}{c}{\texttt{Flare 3}} \\
			1 & $6.99^{+0.81}_{-0.67}$ & $1.40^{+0.08}_{-0.04}$ & $>8.02$ & $2.01^{+0.07}_{-0.05}$ & $2.89^{+0.10}_{-0.08}$ & $0.10^{+0.02}_{-0.01}$ & $1.16^{+0.13}_{-0.11}$ & 1.19 & $14.32^{+0.17}_{-0.17}$ \\
			2 & $7.26^{+0.66}_{-0.57}$ & $1.28^{+0.08}_{-0.08}$ & $7.95^{+1.77}_{-0.99}$ & $1.99^{+0.07}_{-0.05}$ & $2.92^{+0.11}_{-0.09}$ & $0.14^{+0.03}_{-0.02}$ & $1.09^{+0.10}_{-0.09}$ & 1.17 & $20.32^{+0.22}_{-0.23}$ \\
			3 & $7.15^{+0.60}_{-0.55}$ & $1.27^{+0.06}_{-0.08}$ & $8.15^{+1.60}_{-1.02}$ & $2.01^{+0.08}_{-0.05}$ & $2.91^{+0.10}_{-0.09}$ & $0.17^{+0.04}_{-0.02}$ & $1.16^{+0.10}_{-0.09}$ & 1.61 & $23.15^{+0.25}_{-0.25}$ \\
			4 & $7.47^{+0.79}_{-0.64}$ & $1.38^{+0.07}_{-0.08}$ & $8.69^{+0.00}_{-1.46}$ & $2.00^{+0.07}_{-0.04}$ & $2.99^{+0.10}_{-0.08}$ & $0.10^{+0.02}_{-0.01}$ & $1.00^{+0.11}_{-0.09}$ & 1.56 & $15.49^{+0.18}_{-0.18}$ \\
			5 & $6.72^{+0.55}_{-0.48}$ & $1.38^{+0.05}_{-0.04}$ & $>8.60$ & $2.07^{+0.06}_{-0.08}$ & $2.97^{+0.12}_{-0.09}$ & $0.07^{+0.01}_{-0.01}$ & $1.14^{+0.10}_{-0.08}$ & 1.21 & $9.67^{+0.11}_{-0.11}$ \\
			6 & $7.51^{+0.56}_{-0.55}$ & $1.45^{+0.07}_{-0.04}$ & $>8.25$ & $2.20^{+0.06}_{-0.04}$ & $2.84^{+0.08}_{-0.04}$ & $0.09^{+0.02}_{-0.01}$ & $1.02^{+0.09}_{-0.07}$ & 1.42 & $10.79^{+0.12}_{-0.12}$ \\
			7 & $6.62^{+0.50}_{-0.43}$ & $1.35^{+0.06}_{-0.03}$ & $>8.67$ & $2.19^{+0.07}_{-0.05}$ & $2.92^{+0.10}_{-0.07}$ & $0.07^{+0.02}_{-0.01}$ & $1.09^{+0.09}_{-0.08}$ & 1.43 & $9.33^{+0.21}_{-0.21}$ \\
			8 & $5.93^{+0.49}_{-0.42}$ & $1.29^{+0.05}_{-0.04}$ & $>8.83$ & $2.23^{+0.10}_{-0.05}$ & $3.00^{+0.09}_{-0.11}$ & $0.06^{+0.02}_{-0.01}$ & $1.08^{+0.09}_{-0.08}$ & 1.39 & $7.92^{+0.10}_{-0.09}$ \\
			9 & $6.83^{+0.57}_{-0.53}$ & $1.43^{+0.09}_{-0.04}$ & $>8.08$ & $2.32^{+0.06}_{-0.12}$ & $2.85^{+0.12}_{-0.07}$ & $0.08^{+0.02}_{-0.02}$ & $0.92^{+0.08}_{-0.08}$ & 1.37 & $7.97^{+0.10}_{-0.11}$ \\
			10 & $6.44^{+0.49}_{-0.46}$ & $1.25^{+0.06}_{-0.10}$ & $8.78^{+0.00}_{-1.02}$ & $2.26^{+0.11}_{-0.08}$ & $3.00^{+0.14}_{-0.14}$ & $0.07^{+0.03}_{-0.02}$ & $1.02^{+0.08}_{-0.07}$ & 1.06 & $9.19^{+0.11}_{-0.11}$ \\
			11 & $6.49^{+0.44}_{-0.42}$ & $1.37^{+0.09}_{-0.07}$ & $9.24^{+0.00}_{-1.63}$ & $2.07^{+0.07}_{-0.09}$ & $2.96^{+0.14}_{-0.10}$ & $0.08^{+0.02}_{-0.02}$ & $0.99^{+0.07}_{-0.07}$ & 1.09 & $11.63^{+0.13}_{-0.13}$ \\
			12 & $7.61^{+0.91}_{-0.89}$ & $1.32^{+0.06}_{-0.03}$ & $>8.43$ & $2.02^{+0.07}_{-0.05}$ & $2.80^{+0.07}_{-0.07}$ & $0.19^{+0.04}_{-0.03}$ & $1.20^{+0.15}_{-0.10}$ & 1.69 & $33.53^{+0.37}_{-0.37}$ \\
			13 & $7.64^{+0.76}_{-0.67}$ & $1.20^{-0.06}_{+0.06}$ & $8.60^{+1.23}_{-0.92}$ & $2.00^{+0.1}_{-0.04}$ & $2.91^{+0.1}_{-0.0.1}$ & $0.19^{+0.05}_{-0.02}$ & $1.10^{+0.10}_{-0.09}$ & 1.79 & $27.72^{+0.0048}_{-0.0047}$ \\
			\multicolumn{10}{c}{\texttt{Flare 4}} \\
			1 & $6.49^{+0.66}_{-0.57}$ & $1.25^{+0.07}_{-0.08}$ & $>7.77$ & $2.19^{+0.10}_{-0.07}$ & $2.98^{+0.16}_{-0.11}$ & $0.06^{+0.02}_{-0.01}$ & $0.99^{+0.11}_{-0.09}$ & 1.31 & $8.08^{+0.11}_{-0.11}$ \\
			2 & $5.08^{+0.51}_{-0.43}$ & $1.54^{+0.07}_{-0.05}$ & $>8.08$ & $1.96^{+0.06}_{-0.06}$ & $2.88^{+0.11}_{-0.08}$ & $0.03^{+0.004}_{-0.004}$ & $1.05^{+0.11}_{-0.10}$ & 1.02 & $4.37^{+0.10}_{-0.05}$ \\
			3 & $5.28^{+0.49}_{-0.38}$ & $1.71^{+0.11}_{-0.08}$ & $>7.23$ & $1.92^{+0.03}_{-0.03}$ & $2.91^{+0.06}_{-0.05}$ & $0.03^{+0.00}_{-0.00}$ & $1.01^{+0.13}_{-0.11}$ & 1.09 & $6.15^{+0.09}_{-0.08}$ \\
			4 & $4.56^{+0.56}_{-0.48}$ & $1.62^{+0.11}_{-0.07}$ & $>7.16$ & $1.92^{+0.06}_{-0.06}$ & $2.92^{+0.12}_{-0.10}$ & $0.02^{+0.003}_{-0.002}$ & $1.13^{+0.16}_{-0.11}$ & 0.91 & $3.72^{+0.09}_{-0.08}$ \\
			5 & $3.94^{+0.57}_{-0.39}$ & $1.62^{+0.09}_{-0.07}$ & $>8.08$ & $1.79^{+0.08}_{-0.03}$ & $2.99^{+0.05}_{-0.12}$ & $0.02^{+0.00}_{-0.00}$ & $1.03^{+0.13}_{-0.13}$ & 0.96 & $2.63^{+0.06}_{-0.00}$ \\
			6 & $4.23^{+0.55}_{-0.24}$ & $1.52^{+0.09}_{-0.06}$ & $>7.48$ & $1.83^{+0.08}_{-0.04}$ & $2.97^{+0.08}_{-0.12}$ & $0.02^{+0.004}_{-0.002}$ & $1.09^{+0.14}_{-0.12}$ & 1.03 & $3.24^{+0.05}_{-0.04}$ \\
			7 & $5.20^{+0.37}_{-0.34}$ & $1.90^{+0.10}_{-0.09}$ & $>7.79$ & $1.76^{+0.04}_{-0.04}$ & $2.76^{+0.07}_{-0.05}$ & $0.05^{+0.01}_{-0.00}$ & $0.92^{+0.09}_{-0.09}$ & 1.58 & $5.76^{+0.07}_{-0.07}$ \\
			8 & $5.58^{+0.87}_{-0.78}$ & $1.40^{+0.03}_{-0.03}$ & $>9.19$ & $1.88^{+0.02}_{-0.03}$ & $1.99^{+0.10}_{-0.11}$ & $0.17^{+0.02}_{-0.02}$ & $0.91^{+0.12}_{-0.09}$ & 1.08 & $11.15^{+0.13}_{-0.12}$ \\
			9 & $4.36^{+0.77}_{-0.97}$ & $1.28^{+0.03}_{-0.02}$ & $>9.46$ & $1.82^{+0.03}_{-0.04}$ & $2.02^{+0.21}_{-0.08}$ & $0.15^{+0.013}_{-0.026}$ & $1.52^{+0.22}_{-0.13}$ & 1.43 & $11.78^{+0.24}_{-0.12}$ \\
			10 & $4.61^{+0.82}_{-0.48}$ & $1.32^{+0.03}_{-0.03}$ & $>9.05$ & $1.94^{+0.03}_{-0.02}$ & $1.99^{+0.10}_{-0.13}$ & $0.11^{+0.02}_{-0.01}$ & $1.65^{+0.16}_{-0.25}$ & 1.04 & $8.18^{+0.09}_{-0.09}$ \\
			11 & $5.24^{+0.49}_{-0.47}$ & $1.64^{+0.07}_{-0.05}$ & $>3.08$ & $1.83^{+0.06}_{-0.05}$ & $2.93^{+0.08}_{-0.08}$ & $0.03^{+0.005}_{-0.004}$ & $1.02^{+0.10}_{-0.09}$ & 1.39 & $4.97^{+0.06}_{-0.06}$ \\
			\multicolumn{10}{c}{\texttt{Flare 5}} \\
			1 & $5.31_{-0.83}^{+3.73}$ & $1.62_{-0.11}^{+0.1}$ & $>6.122$ & $2.1_{-0.03}^{+0.02}$ & $1.31_{-0.05}^{+0.18}$ & $0.050_{-0.010}^{+0.020}$ & $1.16_{-0.2}^{+0.23}$ & $1.43$ & $2.78_{-0.037}^{+0.037}$ \\
			2 & $4.54_{-0.85}^{+1.36}$ & $1.60_{-0.04}^{+0.04}$ & $>8.726$ & $1.86_{-0.02}^{+0.03}$ & $1.7_{-0.06}^{+0.04}$ & $0.050_{-0.010}^{+0.010}$ & $1.11_{-0.12}^{+0.32}$ & $1.92$ & $3.888_{-0.048}^{+0.048}$ \\
			3 & $3.65_{-0.0}^{+0.0}$ & $1.66_{-0.0}^{+0.0}$ & $10$ & $1.46_{-0.0}^{+0.0}$ & $2.29_{-0.0}^{+0.0}$ & $0.110_{-0.0}^{+0.0}$ & $0.73_{-0.0}^{+0.0}$ & $2.02$ & $9.225_{-0.113}^{+0.116}$ \\
			4 & $24.78_{0.0}^{0.0}$ & $1.47_{-0.0}^{+0.0}$ & $10$ & $1.81_{-0.0}^{+0.0}$ & $1.37_{-0.0}^{+0.0}$ & $0.17_{-0.0}^{+0.0}$ & $0.79_{-0.0}^{+0.0}$ & $2.50$ & $12.61_{-0.14}^{+0.14}$ \\
			5 & $4.55_{-1.22}^{+1.35}$ & $1.54_{-0.05}^{+0.06}$ & $>8.174$ & $1.88_{-0.03}^{+0.03}$ & $1.67_{-0.37}^{+0.19}$ & $0.100_{-0.010}^{+0.010}$ & $0.79_{-0.08}^{+0.1}$ & $1.34$ & $7.106_{-0.082}^{+0.084}$ \\
			6 & $2.86_{-0.61}^{+1.04}$ & $1.99_{-0.14}^{+0.1}$ & $>9$ & $2.09_{-0.03}^{+0.03}$ & $1.3_{-0.06}^{+0.16}$ & $0.100_{-0.020}^{+0.020}$ & $0.68_{-0.09}^{+0.15}$ & $1.50$ & $6.131_{-0.092}^{+0.093}$ \\
			7 & $3.52_{-0.0}^{+0.0}$ & $2.30_{-0.0}^{+0.0}$ & $9.065$ & $2.1_{-0.0}^{+0.0}$ & $1.3_{-0.0}^{+0.0}$ & $0.100_{-0.0}^{+0.0}$ & $0.61_{-0.0}^{+0.0}$ & $2.08$ & $4.535_{-0.073}^{+0.074}$ \\
			\multicolumn{10}{c}{\texttt{NuStar}}	\\	
			1 & $2.41^{+0.80}_{-1.05}$ & $1.74^{+0.02}_{-0.02}$ & $>9.76$ & $1.82^{+0.02}_{-0.02}$ & $1.70^{+0.05}_{-0.05}$ & $0.008^{+0.0001}_{-0.0002}$ & $0.98^{+0.02}_{-0.02}$ & 1.14 & $0.56^{+0.01}_{-0.01}$ \\
			2 & $3.42^{+0.85}_{-0.78}$ & $1.72^{+0.02}_{-0.02}$ & $>9.82$ & $1.87^{+0.02}_{-0.02}$ & $1.73^{+0.09}_{-0.09}$ & $0.010^{+0.0003}_{-0.0004}$ & $0.99^{+0.02}_{-0.02}$ & 1.27 & $0.67^{+0.01}_{-0.02}$ \\
			3 & $4.15^{+0.83}_{-0.78}$ & $1.73^{+0.02}_{-0.02}$ & $>9.71$ & $1.88^{+0.02}_{-0.02}$ & $1.86^{+0.10}_{-0.09}$ & $0.010^{+0.0004}_{-0.0004}$ & $1.00^{+0.02}_{-0.02}$ & 1.09 & $0.76^{+0.02}_{-0.02}$ \\
			4 & $2.95^{+0.80}_{-0.75}$ & $1.71^{+0.02}_{-0.02}$ & $>9.73$ & $1.86^{+0.02}_{-0.02}$ & $1.76^{+0.09}_{-0.13}$ & $0.010^{+0.0004}_{-0.0004}$ & $1.00^{+0.02}_{-0.02}$ & 1.10 & $0.73^{+0.02}_{-0.02}$ \\
			5 & $4.19^{+1.14}_{-0.76}$ & $1.73^{+0.02}_{-0.02}$ & $>9.79$ & $1.95^{+0.02}_{-0.01}$ & $1.27^{+0.27}_{-0.05}$ & $0.016^{+0.0004}_{-0.0004}$ & $0.98^{+0.02}_{-0.01}$ & 1.30 & $1.12^{+0.04}_{-0.05}$ \\
			6 & $3.80^{+0.52}_{-0.51}$ & $1.72^{+0.01}_{-0.01}$ & $>9.77$ & $1.97^{+0.01}_{-0.01}$ & $1.81^{+0.07}_{-0.06}$ & $0.019^{+0.0005}_{-0.0005}$ & $0.96^{+0.01}_{-0.01}$ & 1.24 & $1.35^{+0.03}_{-0.04}$ \\
			7 & $5.51^{+0.50}_{-0.50}$ & $1.70^{+0.01}_{-0.01}$ & $>9.72$ & $2.09^{+0.01}_{-0.02}$ & $2.00^{+0.04}_{-0.03}$ & $0.030^{+0.0008}_{-0.0011}$ & $0.95^{+0.01}_{-0.01}$ & 1.11 & $2.27^{+0.04}_{-0.04}$ \\
			8 & $7.40^{+0.90}_{-0.74}$ & $1.71^{+0.02}_{-0.01}$ & $>9.14$ & $2.07^{+0.01}_{-0.03}$ & $2.01^{+0.10}_{-0.02}$ & $0.030^{+0.0009}_{-0.0021}$ & $1.00^{+0.01}_{-0.01}$ & 1.10 & $1.76^{+0.02}_{-0.02}$ \\
			9 & $7.19^{+0.71}_{-0.69}$ & $1.70^{+0.01}_{-0.02}$ & $>9.81$ & $2.08^{+0.02}_{-0.02}$ & $1.00^{+0.04}_{-0.03}$ & $0.038^{+0.0016}_{-0.0015}$ & $1.00^{+0.01}_{-0.01}$ & 1.25 & $2.47^{+0.03}_{-0.02}$ \\
			10 & $8.38^{+2.34}_{-2.02}$ & $1.67^{+0.03}_{-0.04}$ & $>9.36$ & $1.93^{+0.04}_{-0.04}$ & $1.06^{+0.10}_{-0.10}$ & $0.023^{+0.0016}_{-0.0013}$ & $1.00^{+0.03}_{-0.03}$ & 1.39 & $1.45^{+0.38}_{-0.16}$ \\
		\enddata
	\end{deluxetable*}


			\setlength{\tabcolsep}{4pt} 
		\begin{deluxetable*}{cccccccccccc}
			 \digitalasset
			\tablenum{5} 
		
			\tablecaption{Spectral best-fit parameters for the Model B for \textit{AstroSat} and \textit{NuSTAR} data in the 0.7--30.0 keV and 3.0--79.0 keV energy ranges.The flux reported here is the unabsorbed flux in the energy range 0.7--30 keV  measured in $10^{-9}$ erg cm$^{-2}$ s$^{-1}$. The ionization parameter is given in erg cm s$^{-1}$.}

			\label{Table 4}
			\tablewidth{0pt}
			\tabletypesize{\scriptsize}
\tablehead{
	\colhead{\small \textbf{Seg}} &
	\colhead{\small \textbf{zxipcf}} &
	\colhead{\small \textbf{zxipcf}} &
	\colhead{\small \textbf{diskbb}} &
	\colhead{\small \textbf{xillverCp}} &
	\colhead{\small \textbf{xillverCp}} &
	\colhead{\small \textbf{constant}} &
	\colhead{\small \textbf{thcomp}} &
	\colhead{\small $\boldsymbol{\chi^2_{\rm red}}$} &
	\colhead{\small \textbf{Flux}} &
	\colhead{\small \textbf{Luminosity}} &
	\colhead{\small \textbf{Disk  Luminosity}}\\
	\colhead{} &
	\colhead{\small $N_{\rm H}$ ($10^{22}$ cm$^{-2}$)} &
	\colhead{\small $\log \xi$} &
	\colhead{\small $T_{\rm in}$ (keV)} &
	\colhead{\small $\log \xi$} &
	\colhead{\small Norm} &
	\colhead{\small Factor} &
	\colhead{\small $\Gamma$} &
	\colhead{} &
	\colhead{\small $(10^{-9} \,\text{erg/cm}^2/\text{s})$}&
	\colhead{\small ($10^{37}$ erg s$^{-1}$)} &
	\colhead{\small  ($10^{37}$ erg s$^{-1}$)}
}
        \startdata
		\multicolumn{12}{c}{\texttt{Flare1}}	\\																							
		1	&	$51.37^{+20.54}_{-16.81}$	&	$3.17^{+0.11}_{-0.10}$	&	$0.80^{+0.05}_{-0.02}$	&	$<0.28$	&	$0.04^{+0.01}_{-0.04}$	&	$1.14^{+0.07}_{-0.12}$	&	$2.25^{+0.10}_{-0.10}$	&	1.30	&	$6.56^{+0.70}_{-1.69}$	&	$5.8^{+0.62}_{-1.5}$	&	$2.79^{+0.57}_{-1.08}$	\\
		2	&	$14.81^{+6.18}_{-5.11}$	&	$2.53^{+0.19}_{-0.20}$	&	$0.92^{+0.05}_{-0.05}$	&	$<0.20$	&	$0.09^{+0.04}_{-0.00}$	&	$1.02^{+0.05}_{-0.12}$	&	$2.31^{+0.07}_{-0.07}$	&	1.25	&	$10.62^{+1.77}_{-1.55}$	&	$9.4^{+1.57}_{-1.37}$	&	$5.57^{+1.15}_{-1.81}$	\\
		3	&	$25.51^{+4.46}_{-10.01}$	&	$2.57^{+0.09}_{-0.11}$	&	$1.09^{+0.07}_{-0.08}$	&	$<0.40$	&	$0.10^{+0.02}_{-0.08}$	&	$0.94^{+0.04}_{-0.06}$	&	$2.30^{+0.08}_{-0.17}$	&	1.03	&	$22.96^{+4.39}_{-4.83}$	&	$20.3^{+3.88}_{-4.27}$	&	$7.14^{+0.37}_{-1.44}$	\\
		4	&	$16.49^{+2.78}_{-1.37}$	&	$2.30^{+0.10}_{-0.13}$	&	$1.08^{+0.04}_{-0.02}$	&	$<0.10$	&	$0.12^{+0.04}_{-0.06}$	&	$1.03^{+0.07}_{-0.05}$	&	$2.21^{+0.06}_{-0.09}$	&	1.49	&	$18.73^{+1.55}_{-1.54}$	&	$16.6^{+1.37}_{-1.36}$	&	$6.65^{+1.21}_{-1.82}$	\\
		5	&	$3.66^{+5.23}_{-1.64}$	&	$2.66^{+0.51}_{-0.58}$	&	$1.35^{+0.04}_{-0.04}$	&	$>0.82$	&	$0.27^{+0.07}_{-0.21}$	&	$0.95^{+0.07}_{-0.07}$	&	$2.21^{+0.12}_{-0.05}$	&	1.16	&	$61.19^{+3.79}_{-4.57}$	&	$54.1^{+3.35}_{-4.04}$	&	$24.65^{+1.44}_{-1.58}$	\\
		6	&	$4.66^{+6.44}_{-2.31}$	&	$2.60^{+0.54}_{-0.21}$	&	$1.42^{+0.03}_{-0.03}$	&	$>0.82$	&	$0.32^{+0.09}_{-0.28}$	&	$0.98^{+0.07}_{-0.06}$	&	$2.21^{+0.13}_{-0.04}$	&	1.17	&	$71.58^{+8.74}_{-5.03}$	&	$63.3^{+7.73}_{-4.45}$	&	$41.30^{+6.39}_{-7.35}$	\\
		7	&	$15.99^{+7.29}_{-9.94}$	&	$2.93^{+0.12}_{-0.45}$	&	$1.19^{+0.03}_{-0.05}$	&	$<0.13$	&	$0.17^{+0.08}_{-0.08}$	&	$1.11^{+0.05}_{-0.08}$	&	$2.18^{+0.07}_{-0.08}$	&	1.48	&	$30.19^{+1.01}_{-1.72}$	&	$26.7^{+0.89}_{-1.52}$	&	$8.79^{+3.77}_{-3.76}$	\\
		8	&	$16.44^{+3.15}_{-1.81}$	&	$2.39^{+0.16}_{-0.11}$	&	$1.27^{+0.03}_{-0.03}$	&	$<0.14$	&	$0.18^{+0.05}_{-0.18}$	&	$0.84^{+0.08}_{-0.04}$	&	$2.19^{+0.10}_{-0.06}$	&	1.03	&	$48.07^{+5.31}_{-4.78}$	&	$42.5^{+4.70}_{-4.23}$	&	$15.95^{+3.74}_{-3.06}$	\\
		9	&	$6.71^{+3.17}_{-2.11}$	&	$2.34^{+0.17}_{-0.30}$	&	$1.37^{+0.02}_{-0.02}$	&	$>0.82$	&	$0.30^{+0.11}_{-0.05}$	&	$0.94^{+0.06}_{-0.06}$	&	$2.20^{+0.04}_{-0.08}$	&	1.37	&	$69.35^{+6.93}_{-7.14}$	&	$61.4^{+6.13}_{-6.32}$	&	$33.29^{+4.67}_{-6.56}$	\\
		10	&	$9.33^{+1.12}_{-4.42}$	&	$2.63^{+0.19}_{-0.28}$	&	$1.25^{+0.02}_{-0.02}$	&	$<0.18$	&	$0.28^{+0.13}_{-0.07}$	&	$0.95^{+0.09}_{-0.05}$	&	$2.35^{+0.05}_{-0.12}$	&	1.20	&	$40.56^{+3.29}_{-3.73}$	&	$35.9^{+2.91}_{-3.3}$	&	$19.98^{+0.70}_{-1.82}$	\\
		11	&	$50.41^{+23.64}_{-15.21}$	&	$3.09^{+0.14}_{-0.13}$	&	$1.09^{+0.06}_{-0.06}$	&	$>0.82$	&	$0.08^{+0.08}_{-0.10}$	&	$0.94^{+0.12}_{-0.10}$	&	$2.21^{+0.14}_{-0.20}$	&	0.89	&	$16.70^{+1.54}_{-1.49}$	&	$14.8^{+1.36}_{-1.32}$	&	$8.74^{+4.02}_{-2.19}$	\\
		12	&	$36.80^{+19.50}_{-10.30}$	&	$2.85^{+0.16}_{-0.17}$	&	$1.14^{+0.03}_{-0.04}$	&	$>0.82$	&	$0.21^{+0.06}_{-0.14}$	&	$1.01^{+0.09}_{-0.07}$	&	$2.23^{+0.09}_{-0.07}$	&	1.12	&	$7.57^{+2.15}_{-1.46}$	&	$6.7^{+1.9}_{-1.29}$	&	$3.36^{+1.00}_{-1.30}$	\\
		\multicolumn{12}{c}{\texttt{Flare2}}	\\																						
		1	&	$77.14^{+6.25}_{-12.83}$	&	$3.14^{+0.13}_{-0.05}$	&	$0.81^{+0.02}_{-0.04}$	&	$<0.48$	&	$0.02^{+0.02}_{-0.02}$	&	$0.99^{+0.05}_{-0.11}$	&	$2.26^{+0.17}_{-0.17}$	&	1.24	&	$5.52^{+0.65}_{-1.01}$	&	$4.88^{+0.58}_{-0.89}$	&	$2.36^{+0.67}_{-0.67}$	\\
		2	&	$74.89^{+8.76}_{-19.60}$	&	$3.31^{+0.15}_{-0.08}$	&	$0.77^{+0.01}_{-0.04}$	&	$<0.41$	&	$0.04^{+0.02}_{-0.02}$	&	$1.25^{+0.10}_{-0.12}$	&	$2.34^{+0.11}_{-0.17}$	&	0.92	&	$6.35^{+0.62}_{-1.76}$	&	$5.62^{+0.55}_{-1.56}$	&	$1.43^{+0.24}_{-0.26}$	\\
		3	&	$44.63^{+6.81}_{-4.03}$	&	$2.72^{+0.03}_{-0.06}$	&	$0.75^{+0.03}_{-0.04}$	&	$<1.29$	&	$0.01^{+0.01}_{-0.01}$	&	$0.98^{+0.16}_{-0.13}$	&	$1.97^{+0.15}_{-0.12}$	&	1.62	&	$4.15^{+0.77}_{-0.69}$	&	$3.67^{+0.68}_{-0.61}$	&	$1.27^{+0.12}_{-0.32}$	\\
		4	&	$61.82^{+13.65}_{-19.53}$	&	$3.26^{+0.14}_{-0.13}$	&	$0.77^{+0.03}_{-0.04}$	&	$<0.45$	&	$0.05^{+0.03}_{-0.02}$	&	$1.15^{+0.09}_{-0.09}$	&	$2.25^{+0.12}_{-0.09}$	&	1.70	&	$6.21^{+0.58}_{-0.68}$	&	$5.49^{+0.51}_{-0.60}$	&	$0.92^{+0.27}_{-0.24}$	\\
		5	&	$68.50^{+9.05}_{-12.54}$	&	$3.20^{+0.02}_{-0.07}$	&	$0.79^{+0.02}_{-0.03}$	&	$<0.36$	&	$0.02^{+0.02}_{-0.01}$	&	$1.04^{+0.07}_{-0.08}$	&	$2.11^{+0.12}_{-0.14}$	&	1.14	&	$7.84^{+1.32}_{-1.18}$	&	$6.94^{+1.17}_{-1.04}$	&	$1.63^{+0.05}_{-0.46}$	\\
		6	&	$42.42^{+25.69}_{-17.79}$	&	$3.39^{+0.33}_{-0.19}$	&	$0.73^{+0.04}_{-0.04}$	&	$<0.48$	&	$0.11^{+0.04}_{-0.04}$	&	$1.13^{+0.12}_{-0.11}$	&	$2.30^{+0.07}_{-0.10}$	&	1.42	&	$10.24^{+1.32}_{-1.19}$	&	$9.06^{+1.17}_{-1.05}$	&	$1.09^{+0.33}_{-0.26}$	\\
		7	&	$25.35^{+20.65}_{-15.16}$	&	$3.05^{+0.35}_{-0.63}$	&	$0.79^{+0.03}_{-0.04}$	&	$<0.36$	&	$0.13^{+0.08}_{-0.04}$	&	$1.28^{+0.19}_{-0.16}$	&	$2.24^{+0.10}_{-0.09}$	&	1.21	&	$10.96^{+1.58}_{-1.27}$	&	$9.70^{+1.40}_{-1.12}$	&	$1.81^{+0.66}_{-0.45}$	\\
		8	&	$18.79^{+3.98}_{-3.93}$	&	$2.18^{+0.13}_{-0.12}$	&	$1.14^{+0.05}_{-0.06}$	&	$>0.82$	&	$0.24^{+0.15}_{-0.08}$	&	$1.04^{+0.12}_{-0.10}$	&	$2.26^{+0.10}_{-0.09}$	&	1.50	&	$23.42^{+2.29}_{-2.24}$	&	$20.7^{+2.03}_{-1.98}$	&	$16.97^{+2.32}_{-3.18}$	\\
		9	&	$16.69^{+1.49}_{-1.40}$	&	$2.29^{+0.10}_{-0.05}$	&	$1.36^{+0.02}_{-0.05}$	&	$<0.15$	&	$0.21^{+0.22}_{-0.04}$	&	$0.86^{+0.03}_{-0.05}$	&	$2.22^{+0.07}_{-0.03}$	&	1.34	&	$40.11^{+1.12}_{-2.21}$	&	$35.5^{+0.99}_{-1.96}$	&	$21.87^{+2.02}_{-5.53}$	\\
		10	&	$11.80^{+2.77}_{-2.77}$	&	$2.59^{+0.15}_{-0.15}$	&	$1.40^{+0.01}_{-0.03}$	&	$<0.17$	&	$0.27^{+0.11}_{-0.05}$	&	$0.90^{+0.04}_{-0.04}$	&	$2.19^{+0.06}_{-0.08}$	&	0.87	&	$52.15^{+2.83}_{-2.61}$	&	$46.1^{+2.50}_{-2.31}$	&	$35.69^{+5.38}_{-5.25}$	\\
		11	&	$17.93^{+0.84}_{-0.90}$	&	$1.95^{+0.04}_{-0.03}$	&	$1.39^{+0.03}_{-0.04}$	&	$>0.82$	&	$0.16^{+0.11}_{-0.09}$	&	$0.84^{+0.07}_{-0.06}$	&	$2.12^{+0.10}_{-0.13}$	&	0.84	&	$65.96^{+6.68}_{-5.13}$	&	$58.4^{+5.91}_{-4.54}$	&	$18.71^{+2.00}_{-2.25}$	\\
		12	&	$21.26^{+7.05}_{-5.99}$	&	$2.27^{+0.09}_{-0.15}$	&	$1.07^{+0.03}_{-0.06}$	&	$<0.60$	&	$0.04^{+0.04}_{-0.01}$	&	$1.10^{+0.17}_{-0.14}$	&	$1.96^{+0.16}_{-0.13}$	&	1.87	&	$17.92^{+3.09}_{-1.07}$	&	$15.9^{+2.73}_{-0.95}$	&	$5.73^{+1.35}_{-1.39}$	\\
		13	&	$47.07^{+20.20}_{-15.61}$	&	$3.16^{+0.17}_{-0.14}$	&	$0.71^{+0.06}_{-0.07}$	&	$2.82^{+0.34}_{-0.15}$	&	$0.02^{+0.01}_{-0.01}$	&	$0.99^{+0.13}_{-0.10}$	&	$1.88^{+0.08}_{-0.10}$	&	1.18	&	$7.88^{+1.35}_{-1.18}$	&	$6.97^{+1.19}_{-1.04}$	&	$0.90^{+0.57}_{-0.39}$	\\
		\multicolumn{12}{c}{\texttt{Flare3}}	\\																					
		1	&	$37.22^{+5.37}_{-7.19}$	&	$2.90^{+0.06}_{-0.10}$	&	$1.01^{+0.03}_{-0.04}$	&	$<0.16$	&	$0.13^{+0.09}_{-0.03}$	&	$1.19^{+0.08}_{-0.08}$	&	$2.37^{+0.03}_{-0.12}$	&	$1.30$	&	$18.76^{+3.25}_{-2.55}$	&	$16.6^{+2.88}_{-2.26}$	&	$9.12^{+2.81}_{-1.91}$	\\
		2	&	$18.55^{+6.66}_{-3.27}$	&	$2.33^{+0.10}_{-0.14}$	&	$1.14^{+0.04}_{-0.04}$	&	$0.42^{+0.28}_{-0.32}$	&	$0.12^{+0.09}_{-0.05}$	&	$0.93^{+0.09}_{-0.08}$	&	$2.23^{+0.11}_{-0.10}$	&	$1.22$	&	$22.23^{+0.46}_{-2.29}$	&	$19.7^{+0.42}_{-2.03}$	&	$6.66^{+1.26}_{-1.40}$	\\
		3	&	$29.77^{+5.22}_{-8.97}$	&	$2.36^{+0.04}_{-0.12}$	&	$1.28^{+0.03}_{-0.08}$	&	$<0.54$	&	$0.07^{+0.09}_{-0.02}$	&	$0.87^{+0.08}_{-0.07}$	&	$2.16^{+0.15}_{-0.06}$	&	$1.50$	&	$33.76^{+3.05}_{-4.11}$	&	$29.9^{+2.70}_{-3.64}$	&	$10.48^{+2.99}_{-1.72}$	\\
		4	&	$39.65^{+8.94}_{-7.36}$	&	$2.85^{+0.08}_{-0.09}$	&	$1.01^{+0.94}_{-0.01}$	&	$3.97^{+0.15}_{-0.11}$	&	$0.06^{+0.01}_{-0.03}$	&	$0.96^{+0.05}_{-0.08}$	&	$1.94^{+0.03}_{-0.02}$	&	$1.51$	&	$20.93^{+2.48}_{-2.63}$	&	$18.5^{+2.19}_{-2.33}$	&	$10.55^{+1.65}_{-3.46}$	\\
		5	&	$20.18^{+6.76}_{-3.64}$	&	$2.50^{+0.11}_{-0.13}$	&	$0.88^{+0.04}_{-0.04}$	&	$2.66^{+0.13}_{-0.26}$	&	$0.04^{+0.02}_{-0.01}$	&	$1.07^{+0.10}_{-0.09}$	&	$2.20^{+0.11}_{-0.06}$	&	$1.44$	&	$12.94^{+1.96}_{-1.73}$	&	$11.4^{+1.73}_{-1.53}$	&	$1.86^{+1.10}_{-1.30}$	\\
		6	&	$20.14^{+7.43}_{-3.16}$	&	$2.38^{+0.10}_{-0.10}$	&	$0.91^{+0.03}_{-0.04}$	&	$2.69^{+0.12}_{-0.12}$	&	$0.05^{+0.02}_{-0.02}$	&	$0.95^{+0.09}_{-0.07}$	&	$2.28^{+0.08}_{-0.09}$	&	$1.51$	&	$18.39^{+3.32}_{-2.29}$	&	$16.3^{+2.94}_{-2.03}$	&	$2.66^{+0.72}_{-0.49}$	\\
		7	&	$15.50^{+4.21}_{-3.54}$	&	$2.39^{+0.05}_{-0.13}$	&	$0.90^{+0.02}_{-0.04}$	&	$1.70^{+0.24}_{-0.19}$	&	$0.11^{+0.08}_{-0.04}$	&	$1.02^{+0.07}_{-0.09}$	&	$2.53^{+0.04}_{-0.17}$	&	$1.53$	&	$12.51^{+1.95}_{-1.68}$	&	$11.1^{+1.73}_{-1.49}$	&	$4.21^{+0.78}_{-0.58}$	\\
		8	&	$16.34^{+3.37}_{-3.30}$	&	$2.32^{+0.12}_{-0.10}$	&	$0.85^{+0.04}_{-0.05}$	&	$2.72^{+0.11}_{-0.13}$	&	$0.07^{+0.05}_{-0.03}$	&	$0.96^{+0.11}_{-0.09}$	&	$2.48^{+0.10}_{-0.11}$	&	$1.62$	&	$17.19^{+4.32}_{-2.90}$	&	$15.2^{+3.82}_{-2.57}$	&	$2.71^{+0.69}_{-0.59}$	\\
		9	&	$40.59^{+3.85}_{-8.06}$	&	$2.98^{+0.06}_{-0.04}$	&	$0.91^{+0.01}_{-0.03}$	&	$<0.30$	&	$0.10^{+0.08}_{-0.08}$	&	$0.92^{+0.05}_{-0.05}$	&	$2.61^{+0.09}_{-0.11}$	&	$1.48$	&	$8.34^{+0.54}_{-0.34}$	&	$7.38^{+0.48}_{-0.30}$	&	$4.20^{+0.51}_{-0.67}$	\\
		10	&	$45.83^{+9.97}_{-8.23}$	&	$2.96^{+0.08}_{-0.07}$	&	$0.89^{+0.06}_{-0.07}$	&	$>3.66$	&	$0.01^{+0.01}_{-0.01}$	&	$0.96^{+0.08}_{-0.07}$	&	$2.27^{+0.04}_{-0.04}$	&	$1.08$	&	$12.93^{+1.98}_{-1.98}$	&	$11.4^{+1.75}_{-1.75}$	&	$4.20^{+1.15}_{-1.17}$	\\
		11	&	$42.46^{+9.77}_{-7.77}$	&	$3.02^{+0.09}_{-0.08}$	&	$0.98^{+0.03}_{-0.03}$	&	$>0.82$	&	$0.08^{+0.06}_{-0.05}$	&	$1.02^{+0.08}_{-0.07}$	&	$2.39^{+0.11}_{-0.18}$	&	$1.22$	&	$10.78^{+1.11}_{-1.02}$	&	$9.54^{+0.98}_{-0.90}$	&	$6.03^{+0.80}_{-0.71}$	\\
		12	&	$25.91^{+3.19}_{-6.37}$	&	$2.25^{+0.04}_{-0.07}$	&	$1.22^{+0.09}_{-0.06}$	&	$<0.30$	&	$0.14^{+0.10}_{-0.08}$	&	$0.94^{+0.10}_{-0.08}$	&	$2.21^{+0.15}_{-0.12}$	&	$1.62$	&	$26.88^{+1.28}_{-1.11}$	&	$23.8^{+1.13}_{-0.98}$	&	$8.62^{+1.19}_{-0.81}$	\\
		13	&	$19.40^{+1.38}_{-0.76}$	&	$2.18^{+0.06}_{-0.04}$	&	$1.27^{+0.03}_{-0.03}$	&	$<0.18$	&	$0.14^{+0.17}_{-0.06}$	&	$0.98^{+0.09}_{-0.08}$	&	$2.27^{+0.14}_{-0.10}$	&	$1.71$	&	$32.16^{+2.15}_{-3.29}$	&	$28.5^{+1.90}_{-2.91}$	&	$10.71^{+0.96}_{-1.95}$	\\
	\enddata
\end{deluxetable*}
			\setlength{\tabcolsep}{2pt}
		\begin{deluxetable*}{ccccccccccccc}
			\digitalasset
			\tablenum{5}
			
			\tablecaption{Spectral best-fit parameters for the Model B for \textit{AstroSat} and \textit{NuSTAR} data in the 0.7--30.0 keV and 3.0--79.0 keV energy ranges.The flux reported here is the unabsorbed flux in the energy range 0.7--30 keV  measured in $10^{-9}$ erg cm$^{-2}$ s$^{-1}$. The ionization parameter is given in erg cm s$^{-1}$.}

			\label{Table 5}
			\tablewidth{0pt}
			\tabletypesize{\scriptsize}
			\tablehead{
				\colhead{\small \textbf{Seg}} &
				\colhead{\small \textbf{zxipcf}} &
				\colhead{\small \textbf{zxipcf}} &
				\colhead{\small \textbf{diskbb}} &
				\colhead{\small \textbf{xillverCp}} &
				\colhead{\small \textbf{xillverCp}} &
				\colhead{\small \textbf{constant}} &
				\colhead{\small \textbf{thcomp}} &
				\colhead{\small $\boldsymbol{\chi^2_{\rm red}}$} &
				\colhead{\small \textbf{Flux}} &
				\colhead{\small \textbf{Luminosity}} &
				\colhead{\small \textbf{Disk  Luminosity}}\\
				\colhead{} &
				\colhead{\small $N_{\rm H}$ ($10^{22}$ cm$^{-2}$)} &
				\colhead{\small $\log \xi$} &
				\colhead{\small $T_{\rm in}$ (keV)} &
				\colhead{\small $\log \xi$} &
				\colhead{\small Norm} &
				\colhead{\small Factor} &
				\colhead{\small $\Gamma$} &
				\colhead{} &
				\colhead{\small $(10^{-9} \,\text{erg/cm}^2/\text{s})$}&
				\colhead{\small ($10^{37}$ erg s$^{-1}$)} &
				\colhead{\small  ($10^{37}$ erg s$^{-1}$)}
			}
		\startdata
	\multicolumn{12}{c}{\texttt{Flare4}}	\\																						
	1	&	$34.49^{+4.85}_{-5.42}$	&	$3.02^{+0.09}_{-0.07}$	&	$0.91^{+0.03}_{-0.03}$	&	$<1.28$	&	$0.03^{+0.05}_{-0.03}$	&	$1.04^{+0.07}_{-0.05}$	&	$2.42^{+0.18}_{-0.22}$	&	1.33	&	$9.91^{+2.45}_{-1.51}$	&	$8.77^{+2.17}_{-1.34}$	&	$1.94^{+0.36}_{-0.32}$	\\
	2	&	$68.81^{+5.64}_{-15.73}$	&	$3.14^{+0.11}_{-0.05}$	&	$0.79^{+0.01}_{-0.04}$	&	$<0.46$	&	$0.02^{+0.01}_{-0.01}$	&	$1.06^{+0.08}_{-0.10}$	&	$2.26^{+0.11}_{-0.14}$	&	1.19	&	$7.34^{+0.77}_{-1.14}$	&	$6.49^{+0.68}_{-1.01}$	&	$1.75^{+0.30}_{-0.46}$	\\
	3	&	$63.08^{+15.86}_{-19.07}$	&	$3.10^{+0.12}_{-0.09}$	&	$0.53^{+0.06}_{-0.05}$	&	$3.65^{+0.26}_{-0.27}$	&	$0.01^{+0.01}_{-0.00}$	&	$0.88^{+0.10}_{-0.09}$	&	$1.86^{+0.05}_{-0.05}$	&	1.00	&	$6.73^{+1.35}_{-1.23}$	&	$5.95^{+1.19}_{-1.09}$	&	$1.03^{+0.39}_{-0.42}$	\\
	4	&	$44.37^{+12.01}_{-17.85}$	&	$3.34^{+0.26}_{-0.22}$	&	$0.59^{+0.06}_{-0.06}$	&	$2.72^{+0.19}_{-0.24}$	&	$0.01^{+0.01}_{-0.00}$	&	$1.12^{+0.16}_{-0.08}$	&	$2.01^{+0.09}_{-0.08}$	&	1.08	&	$4.59^{+1.69}_{-0.61}$	&	$4.06^{+1.50}_{-0.54}$	&	$1.36^{+0.40}_{-0.72}$	\\
	5	&	$61.53^{+15.30}_{-24.82}$	&	$3.37^{+0.28}_{-0.20}$	&	$0.58^{+0.06}_{-0.06}$	&	$2.69^{+0.18}_{-0.32}$	&	$0.01^{+0.01}_{-0.00}$	&	$0.97^{+0.12}_{-0.12}$	&	$1.94^{+0.07}_{-0.08}$	&	1.00	&	$9.45^{+10.81}_{-3.51}$	&	$8.36^{+9.56}_{-3.11}$	&	$0.78^{+0.37}_{-0.24}$	\\
	6	&	$74.43^{+6.48}_{-12.08}$	&	$3.19^{+0.11}_{-0.06}$	&	$0.74^{+0.03}_{-0.03}$	&	$<0.45$	&	$0.01^{+0.01}_{-0.01}$	&	$1.05^{+0.07}_{-0.09}$	&	$2.11^{+0.15}_{-0.10}$	&	1.38	&	$5.78^{+0.79}_{-1.05}$	&	$5.11^{+0.70}_{-0.93}$	&	$1.31^{+0.21}_{-0.25}$	\\
	7	&	$34.51^{+11.84}_{-9.67}$	&	$3.15^{+0.15}_{-0.14}$	&	$0.61^{+0.07}_{-0.06}$	&	$2.68^{+0.04}_{-0.17}$	&	$0.03^{+0.01}_{-0.00}$	&	$0.94^{+0.09}_{-0.09}$	&	$1.79^{+0.04}_{-0.02}$	&	1.71	&	$6.62^{+0.59}_{-1.10}$	&	$5.86^{+0.52}_{-0.97}$	&	$1.97^{+0.48}_{-0.46}$	\\
	8	&	$32.36^{+5.45}_{-3.97}$	&	$2.93^{+0.16}_{-0.16}$	&	$0.76^{+0.03}_{-0.03}$	&	$<0.06$	&	$0.18^{+0.06}_{-0.05}$	&	$0.93^{+0.13}_{-0.11}$	&	$2.13^{+0.05}_{-0.08}$	&	1.52	&	$18.07^{+2.05}_{-2.22}$	&	$16.00^{+1.81}_{-1.96}$	&	$1.24^{+0.26}_{-0.23}$	\\
	9	&	$20.02^{+7.84}_{-9.79}$	&	$3.12^{+0.19}_{-0.15}$	&	$0.64^{+0.05}_{-0.05}$	&	$<0.09$	&	$0.20^{+0.05}_{-0.04}$	&	$1.18^{+0.14}_{-0.14}$	&	$2.07^{+0.04}_{-0.06}$	&	1.76	&	$19.72^{+1.98}_{-3.63}$	&	$17.4^{+1.75}_{-3.21}$	&	$0.64^{+0.60}_{-0.20}$	\\
	10	&	$27.66^{+6.41}_{-15.56}$	&	$3.27^{+0.37}_{-0.28}$	&	$0.62^{+0.03}_{-0.06}$	&	$<0.16$	&	$0.17^{+0.03}_{-0.04}$	&	$1.27^{+0.18}_{-0.18}$	&	$2.19^{+0.03}_{-0.05}$	&	1.60	&	$12.11^{+1.84}_{-1.38}$	&	$10.7^{+1.63}_{-1.22}$	&	$0.58^{+0.16}_{-0.19}$	\\
	11	&	$73.40^{+9.68}_{-9.38}$	&	$3.05^{+0.06}_{-0.05}$	&	$0.80^{+0.04}_{-0.04}$	&	$2.68^{+0.33}_{-0.82}$	&	$0.01^{+0.00}_{-0.01}$	&	$1.00^{+0.10}_{-0.08}$	&	$1.95^{+0.07}_{-0.04}$	&	1.42	&	$7.96^{+0.40}_{-1.07}$	&	$7.04^{+0.36}_{-0.95}$	&	$1.42^{+0.45}_{-0.55}$	\\
	\multicolumn{12}{c}{\texttt{Flare	5}}	\\																																
1	&	$92.62_{-73.05}^{+167.57}$	&	$3.48_{-3.35}^{+3.93}$	&	$0.49_{-0.45}^{+0.52}$	&	$0.30_{-0.12}^{+0.47}$	&	$0.04_{-0.03}^{+0.05}$	&	$1.56_{-1.28}^{+1.91}$	&	$2.20_{-2.15}^{+2.29}$	&	$1.19$	&	$	3.33	_{-	0.5	}^{+	0.77	}$			&$	2.95	_{-	0.44	}^{+	0.68	}$	&	$0.184^{-0.054}_{+0.069}$\\
2	&	$85.69_{-71.40}^{+102.49}$	&	$3.12_{-3.02}^{+3.23}$	&	$0.58_{-0.54}^{+0.62}$	&	$0.28_{-0.09}^{+0.42}$	&	$0.04_{-0.03}^{+0.05}$	&	$1.59_{-1.34}^{+1.91}$	&	$2.15_{-2.09}^{+2.19}$	&	$1.40$	&	$	4.97	_{-	0.83	}^{+	0.88	}$			&$	4.4	_{-	0.73	}^{+	0.78	}$	&	$0.314^{-0.084}_{+0.123}$\\
3	&	$74.11_{-63.76}^{+82.80}$	&	$2.92_{-2.86}^{+3.01}$	&	$0.62_{-0.57}^{+0.65}$	&	$<0.40$	&	$0.05_{-0.04}^{+0.06}$	&	$0.84_{-0.69}^{+0.97}$	&	$1.80_{-1.77}^{+1.87}$	&	$1.20$	&	$	7.45	_{-	1.2	}^{+	1.32	}$			&$	6.59	_{-	1.06	}^{+	1.17	}$	&	$0.551^{-0.145}_{+0.108}$\\
4	&	$34.40_{-27.88}^{+43.84}$	&	$2.30_{-2.09}^{+2.51}$	&	$0.57_{-0.41}^{+0.74}$	&	$<0.35$	&	$0.13_{-0.10}^{+0.16}$	&	$0.81_{-0.68}^{+0.95}$	&	$1.91_{-1.86}^{+1.96}$	&	$1.22$	&	$	11.48	_{-	1.29	}^{+	2.37	}$			&$	10.2	_{-	1.14	}^{+	2.1	}$	&	$0.467^{-0.337}_{+0.843}$\\
5	&	$47.23_{-35.76}^{+65.29}$	&	$2.92_{-2.78}^{+3.08}$	&	$0.70_{-0.66}^{+0.74}$	&	$<0.34$	&	$0.08_{-0.06}^{+0.11}$	&	$0.89_{-0.78}^{+1.03}$	&	$2.10_{-2.05}^{+2.15}$	&	$1.73$	&	$	9.3	_{-	1.17	}^{+	1.23	}$			&$	8.23	_{-	1.04	}^{+	1.09	}$	&	$0.858^{-0.207}_{+0.230}$\\
6	&	$93.33_{-67.34}^{+121.7}$	&	$3.39_{-3.22}^{+3.65}$	&	$0.64_{-0.60}^{+0.69}$	&	$0.44_{-0.18}^{+0.71}$	&	$0.06_{-0.04}^{+0.11}$	&	$0.94_{-0.78}^{+1.13}$	&	$2.18_{-2.06}^{+2.32}$	&	$1.51$	&	$	7.01	_{-	1.35	}^{+	1.68	}$			&$	6.2	_{-	1.19	}^{+	1.49	}$	&	$0.620^{-0.199}_{+0.292}$\\
7	&	$86.87_{-73.06}^{+92.83}$	&	$2.95_{-2.92}^{+2.98}$	&	$0.78_{-0.72}^{+0.81}$	&	$<0.31$	&	$0.02_{-0.01}^{+0.03}$	&	$0.99_{-0.88}^{+1.19}$	&	$2.09_{-1.99}^{+2.18}$	&	$1.84$	&	$	5.35	_{-	1	}^{+	0.83	}$			&$	4.73	_{-	0.88	}^{+	0.73	}$	&	$1.432^{-0.291}_{+0.085}$\\
\multicolumn{12}{c}{\texttt{NuStar}}	\\																																		
1	&	$17.48^{+3.29}_{-2.96}$	&	$2.31^{+0.10}_{-0.11}$	&	$0.33^{+0.01}_{-0.02}$	&	$2.55^{+0.11}_{-0.10}$	&	$0.004^{+0.0002}_{-0.0002}$	&	$1.00^{+0.02}_{-0.01}$	&	$1.70^{+0.02}_{-0.02}$	&	1.19	&	$0.75^{+0.02}_{-0.02}$	&			$0.66^{+0.02}_{-0.02}$	&	$0.051^{-0.013}_{+0.010}$\\											
2	&	$17.05^{+2.65}_{-2.31}$	&	$2.11^{+0.11}_{-0.12}$	&	$0.39^{+0.01}_{-0.01}$	&	$2.12^{+0.15}_{-0.12}$	&	$0.004^{+0.0002}_{-0.0002}$	&	$0.97^{+0.01}_{-0.01}$	&	$1.78^{+0.03}_{-0.03}$	&	1.27	&	$0.91^{+0.02}_{-0.02}$	&			$0.80^{+0.01}_{-0.01}$	&	$0.106^{-0.008}_{+0.008}$\\											
3	&	$18.87^{+2.70}_{-2.60}$	&	$2.21^{+0.09}_{-0.12}$	&	$0.40^{+0.01}_{-0.01}$	&	$2.10^{+0.10}_{-0.11}$	&	$0.005^{+0.0002}_{-0.0002}$	&	$0.98^{+0.01}_{-0.01}$	&	$1.79^{+0.02}_{-0.02}$	&	1.16	&	$1.01^{+0.03}_{-0.03}$	&			$0.89^{+0.02}_{-0.02}$	&	$0.052^{-0.008}_{+0.017}$\\											
4	&	$30.58^{+3.09}_{-3.06}$	&	$2.31^{+0.05}_{-0.05}$	&	$0.41^{+0.02}_{-0.02}$	&	$2.52^{+0.11}_{-0.10}$	&	$0.004^{+0.0002}_{-0.0002}$	&	$1.00^{+0.01}_{-0.01}$	&	$1.76^{+0.03}_{-0.03}$	&	1.25	&	$1.16^{+0.05}_{-0.04}$	&			$1.02^{+0.04}_{-0.03}$	&	$0.077^{-0.013}_{+0.013}$\\										
5	&	$34.20^{+2.58}_{-2.77}$	&	$2.20^{+0.03}_{-0.03}$	&	$0.49^{+0.02}_{-0.02}$	&	$2.46^{+0.10}_{-0.09}$	&	$0.006^{+0.0004}_{-0.0003}$	&	$0.98^{+0.01}_{-0.01}$	&	$1.84^{+0.03}_{-0.03}$	&	1.43	&	$1.38^{+0.03}_{-0.02}$	&			$1.22^{+0.02}_{-0.02}$	&	$0.090^{-0.020}_{+0.020}$	\\										
6	&	$9.95^{+2.07}_{-2.99}$	&	$2.56^{+0.21}_{-0.22}$	&	$0.44^{+0.02}_{-0.02}$	&	$0.66^{+0.04}_{-0.04}$	&	$0.013^{+0.0007}_{-0.0005}$	&	$0.97^{+0.01}_{-0.01}$	&	$2.00^{+0.02}_{-0.02}$	&	1.31	&	$1.84^{+0.03}_{-0.03}$	&			$1.63^{+0.02}_{-0.02}$	&	$0.131^{-0.012}_{+0.022}$	\\										
7	&	$35.16^{+1.17}_{-2.35}$	&	$2.12^{+0.02}_{-0.02}$	&	$0.67^{+0.01}_{-0.01}$	&	$2.00^{+0.04}_{-0.04}$	&	$0.009^{+0.0006}_{-0.0007}$	&	$0.98^{+0.01}_{-0.01}$	&	$2.02^{+0.01}_{-0.02}$	&	1.56	&	$2.32^{+0.09}_{-0.09}$	&			$2.05^{+0.08}_{-0.08}$	&	$0.237^{-0.039}_{+0.038}$	\\										
8	&	$14.41^{+2.24}_{-2.14}$	&	$2.32^{+0.20}_{-0.15}$	&	$0.52^{+0.02}_{-0.03}$	&	$0.31^{+0.04}_{-0.04}$	&	$0.026^{+0.001}_{-0.001}$	&	$0.98^{+0.01}_{-0.01}$	&	$2.16^{+0.01}_{-0.02}$	&	1.15	&	$2.33^{+0.10}_{-0.11}$	&			$2.06^{+0.08}_{-0.09}$	&	$0.241^{-0.049}_{+0.045}$	\\										
9	&	$47.66^{+9.51}_{-8.14}$	&	$3.19^{+0.07}_{-0.08}$	&	$0.43^{+0.01}_{-0.02}$	&	$0.15^{+0.05}_{-0.05}$	&	$0.037^{+0.002}_{-0.001}$	&	$0.99^{+0.01}_{-0.01}$	&	$2.19^{+0.02}_{-0.01}$	&	1.23	&	$2.42^{+0.06}_{-0.05}$	&			$2.14^{+0.05}_{-0.04}$	&	$0.149^{-0.046}_{+0.062}$	\\										
10	&	$35.16^{+15.24}_{-7.89}$	&	$3.02^{+0.19}_{-0.26}$	&	$0.39^{+0.03}_{-0.03}$	&	$0.33^{+0.11}_{-0.13}$	&	$0.019^{+0.002}_{-0.001}$	&	$0.97^{+0.03}_{-0.03}$	&	$2.01^{+0.06}_{-0.04}$	&	1.32	&	$1.74^{+0.09}_{-0.09}$	&			$1.54^{+0.08}_{-0.07}$	&	$0.083^{-0.022}_{+0.027}$	\\										
	\enddata
\end{deluxetable*}

\setlength{\tabcolsep}{32pt}
\begin{deluxetable*}{ccccc}
	\digitalasset
	\tablenum{6}
	\tablecaption{
		Broken power-law fit results for Model A (\texttt{constant}~$\times$~\texttt{TBabs}~$\times$~\texttt{relxillCp}). 
		The best-fit parameters are derived using the broken power-law function: 
		\ensuremath{
			f(x) = 
			\begin{cases} 
				A \, x^{p_1}, & x < x_{\text{break}} \\
				A \, x_{\text{break}}^{p_1 + p_2} \, x^{-p_2}, & x \geq x_{\text{break}}
			\end{cases}
		}
		where \ensuremath{A} is the normalization constant, and \ensuremath{p_1}, \ensuremath{p_2} are the power-law indices before and after the break flux \ensuremath{x_{\text{break}}}, respectively. 
		The parameters listed here describe how each spectral parameter varies with unabsorbed flux in the 0.7–30~keV energy range (in units of \ensuremath{10^{-9}} erg\,cm\ensuremath{^{-2}}\,s\ensuremath{^{-1}}).
	}
	\label{tab:broken_power_law_fits}
	\tablewidth{0pt}
	\tabletypesize{\small}
	\tablehead{
		\colhead{\textbf{Parameter}} & 
		\colhead{\textbf{\ensuremath{A}}} & 
		\colhead{\textbf{\ensuremath{p_1}}} & 
		\colhead{\textbf{\ensuremath{p_2}}} & 
		\colhead{\textbf{\ensuremath{x_{\text{break}}}}}
	}
	\startdata
	Relxillcp \ensuremath{\Gamma} & \ensuremath{1.92 \pm 0.05} & \ensuremath{0.08 \pm 0.07} & \ensuremath{0.00 \pm 0.01} & \ensuremath{1.76 \pm 1.29} \\
	Tbabs \ensuremath{N_{\mathrm{H}}} & \ensuremath{4.34 \pm 0.55} & \ensuremath{0.93 \pm 0.39} & \ensuremath{0.00 \pm 0.03} & \ensuremath{1.45 \pm 0.27} \\
	Relxillcp \ensuremath{\log \xi} & \ensuremath{1.61 \pm 0.15} & \ensuremath{0.36 \pm 0.09} & \ensuremath{0.00 \pm 0.03} & \ensuremath{4.53 \pm 1.00} \\
	Relxillcp \ensuremath{R_{\mathrm{in}}} & \ensuremath{1.71 \pm 0.03} & \ensuremath{0.00 \pm 0.04} & \ensuremath{0.13 \pm 0.01} & \ensuremath{2.31 \pm 0.67} \\
	\enddata
\end{deluxetable*}

\begin{deluxetable*}{ccccc}
	\digitalasset
	\tablenum{7}
	\tablecaption{
		Broken power-law fit results for Model B (\texttt{constant}~$\times$~\texttt{TBabs}~$\times$~\texttt{zxipcf(thcomp*diskbb + xillverCp)}). 
		The best-fit parameters are derived using the broken power-law function:
		\ensuremath{
			f(x) = 
			\begin{cases} 
				A \, x^{p_1}, & x < x_{\text{break}} \\
				A \, x_{\text{break}}^{p_1 + p_2} \, x^{-p_2}, & x \geq x_{\text{break}}
			\end{cases}
		}
		where \ensuremath{A} is the normalization constant, and \ensuremath{p_1}, \ensuremath{p_2} are the power-law indices before and after the break flux \ensuremath{x_{\text{break}}}, respectively.
	}
	\label{tab:broken_power_law_fits_B}
	\tablewidth{0pt}
	\tabletypesize{\small}
	\tablehead{
		\colhead{\textbf{Parameter}} & 
		\colhead{\textbf{\ensuremath{A}}} & 
		\colhead{\textbf{\ensuremath{p_1}}} & 
		\colhead{\textbf{\ensuremath{p_2}}} & 
		\colhead{\textbf{\ensuremath{x_{\text{break}}}}}
	}
	\startdata
	Zxipcf \ensuremath{\log \xi} & \ensuremath{2.28 \pm 0.12} & \ensuremath{0.20 \pm 0.04} & \ensuremath{0.14 \pm 0.02} & \ensuremath{5.67 \pm 0.83} \\
	Thcomp \ensuremath{\Gamma} & \ensuremath{1.85 \pm 0.06} & \ensuremath{0.07 \pm 0.02} & \ensuremath{0.01 \pm 0.03} & \ensuremath{12.51 \pm 4.54} \\
	Zxipcf \ensuremath{N_{\mathrm{H}}} & \ensuremath{18.75 \pm 4.51} & \ensuremath{0.85 \pm 0.18} & \ensuremath{0.90 \pm 0.12} & \ensuremath{5.38 \pm 0.40} \\
	Diskbb \ensuremath{T_{\mathrm{in}}} & \ensuremath{0.38 \pm 0.02} & \ensuremath{0.33 \pm 0.02} & \ensuremath{0.00 \pm 0.32} & \ensuremath{51.89 \pm 12.88} \\
	\enddata
\end{deluxetable*}
\begin{deluxetable*}{lcccccc}
	\tablenum{8}
	\tablecaption{Results of the F-test comparing single power-law (PL) and broken power-law (BPL) fits for different parameters vs flux for Model A.}
	\label{tab:f_test_model1}
	\tablewidth{0pt}
	\tabletypesize{\small}
	\tablehead{
		\colhead{Parameter} & 
		\colhead{$\chi^2_{\rm PL}$} & 
		\colhead{$\chi^2_{\rm BPL}$} & 
		\colhead{F} & 
		\colhead{p-value} & 
		\colhead{$\chi^2_{\mathrm{PL}}$/dof} & 
		\colhead{$\chi^2_{\mathrm{BPL}}$/dof}
	}
	\startdata
	Relxillcp \ensuremath{\Gamma}         & 0.44  & 0.49  & -3.11 & $1.0$    & 0.44/61 & 0.49/59 \\
	Tbabs \ensuremath{N_{\mathrm{H}}} & 14.93 & 18.17 & -5.26 & $1.0$    & 14.93/61 & 18.17/59 \\
	Relxillcp \ensuremath{R_{\mathrm{in}}}& 0.44  & 0.35  & 8.28  & $7.0 \times 10^{-4}$   &0.44/61 & 0.35/59 \\
	Relxillcp \ensuremath{\log \xi}       & 6.05  & 5.07  & 5.70  & $5.5 \times 10^{-3}$   & 6.05/61 & 5.07/59 \\
	\enddata
\end{deluxetable*}

\begin{deluxetable*}{lcccccc}
	\tablenum{9}
	\tablecaption{Results of the F-test comparing single power-law (PL) and broken power-law (BPL) fits for different parameters vs flux for Model B.}
	\label{tab:f_test_model2}
	\tablewidth{0pt}
	\tabletypesize{\small}
	\tablehead{
		\colhead{Parameter} & 
		\colhead{$\chi^2_{\rm PL}$} & 
		\colhead{$\chi^2_{\rm BPL}$} & 
		\colhead{F} & 
		\colhead{p-value} & 
		\colhead{$\chi^2_{\mathrm{PL}}$/dof} &
		\colhead{$\chi^2_{\mathrm{BPL}}$/dof}
	}
	\startdata
Zxipcf \ensuremath{\log \xi}    & 4.277   & 2.078    & 32.813 & $1.9\times10^{-10}$   & 4.227/64 &  2.078/62  \\
Thcomp \ensuremath{\Gamma}       & 0.843   & 0.751    & 3.786  & $2.8\times10^{-2}$     & 0.843/64 & 0.751/62  \\
Zxipcf \ensuremath{N_{\mathrm{H}}}    & 1436.8 & 400.30     & 80.272 & $1.1\times10^{-16}$  & 1436.8/64 & 400.30/62 \\
Diskbb \ensuremath{T_{\mathrm{in}}}  & 1.120   & 1.102   & 0.506  & $6.1\times10^{-1}$      & 1.120/64 & 1.102/62 \\
	\enddata
\end{deluxetable*}

\end{document}